\documentclass[12pt,preprint,referee]{aastex}

\RequirePackage{lineno}
   \usepackage{color}
\usepackage{txfonts}

\usepackage{url}
\usepackage{gensymb}

\newcommand{\be} {\begin{equation}}

\newcommand{\CXO}{{\it Chandra}\,}

\newcommand{\xmm}{{\em XMM-Newton}}
\newcommand{\XMM}{{\em XMM-Newton}}

\newcommand{\fermi}{{\em Fermi}}

\newcommand{\bc}{\begin{center}}
\newcommand{\ec}{\end{center}}
\def\ltsima{$\; \buildrel < \over \sim \;$}
\def\lsim{\lower.5ex\hbox{\ltsima}}
\def\loe{\lower.5ex\hbox{\ltsima}}
\def\gtsima{$\; \buildrel > \over \sim \;$}
\def\gsim{\lower.5ex\hbox{\gtsima}}
\def\goe{\lower.5ex\hbox{\gtsima}}

\def\ltsima{$\; \buildrel < \over \sim \;$}
\def\lsim{\lower.5ex\hbox{\ltsima}}
\def\loe{\lower.5ex\hbox{\ltsima}}
\def\gtsima{$\; \buildrel > \over \sim \;$}
\def\gsim{\lower.5ex\hbox{\gtsima}}
\def\goe{\lower.5ex\hbox{\gtsima}}

\def\ergscm2 {erg\,s$^{-1}$cm$^{-2}$}

\def\cm2 {cm$^{-2}$}

\def\psrj {PSR\,J1907$+$0602}
\def\psrjj {PSR\,J1907$+$0631}
\def\fgl {4FGL\,J1906.2$+$0631}

\def\mgro {MGRO~J1908+06}
\def\snr {SNR~G40.5$-$0.5}
\def\fermi {\emph{Fermi}-LAT}
\def\nsource {Fermi J1906+0626}

\newcommand{\twCO}{$^{12}$CO}
\newcommand{\thCO}{$^{13}$CO}
\newcommand{\CeiO}{C$^{18}$O}
\newcommand{\otz}{$J$=\,1--0}
\newcommand{\km}{\,{\rm km}}

\newcommand{\ps}{\,{\rm s}^{-1}}
\newcommand{\kpc}{\,{\rm kpc}}

\newcommand{\NHH}{N({\rm H}_2)}

\usepackage{color}

\shortauthors{Jian Li, Ruo-Yu Liu, Emma de O\~na Wilhelmi et al.}
\shorttitle{MGRO}

\begin{document}
\title{Investigating the nature of \mgro\/ with multiwavelength observations}

\author{Jian Li\altaffilmark{1\dag,2}, Ruo-Yu Liu\altaffilmark{3\ddag}, Emma de O\~na Wilhelmi\altaffilmark{4\S}, Diego F. Torres$^{5,6,7}$, Qian-Cheng Liu$^{3}$, Matthew Kerr$^{8}$, Rolf B$\ddot{u}$hler$^{4}$, Yang Su$^{9}$, Hao-Ning He$^{10}$, Meng-Yuan Xiao$^{3}$}

\altaffiltext{1}{CAS Key Laboratory for Research in Galaxies and Cosmology, Department of Astronomy, University of Science and Technology of China, Hefei 230026, China $^\dag$jianli@ustc.edu.cn}
\altaffiltext{2}{School of Astronomy and Space Science, University of Science and Technology of China, Hefei 230026, China}
\altaffiltext{3}{Department of Astronomy, Nanjing University, 163 Xianlin Avenue, Nanjing 210023, China $^\ddag$ryliu@nju.edu.cn}
\altaffiltext{4}{Deutsches Elektronen-Synchrotron DESY, D-15738 Zeuthen, Germany $^\S$emma.de.ona.wilhelmi@desy.de}
\altaffiltext{5}{Institute of Space Sciences (ICE, CSIC), Campus UAB, Carrer de Can Magrans, 08193, Barcelona, Spain}
\altaffiltext{6}{Institut d'Estudis Espacials de Catalunya (IEEC), 08034 Barcelona, Spain}
\altaffiltext{7}{Instituci\'o Catalana de Recerca i Estudis Avan\c{c}ats (ICREA), E-08010, Barcelona, Spain}
\altaffiltext{8}{Space Science Division, Naval Research Laboratory, Washington, DC 20375, USA}
\altaffiltext{9}{Purple Mountain Observatory and Key Laboratory of Radio Astronomy, Chinese Academy of Sciences, Nanjing 210034, China}
\altaffiltext{10}{Key Laboratory of Dark Matter and Space Astronomy, Purple Mountain Observatory, Chinese Academy of Sciences, 210023 Nanjing, Jiangsu, China}
\begin{abstract}
The unidentified TeV source \mgro, with emission extending from hundreds of GeV to beyond 100\,TeV, is one of the most intriguing sources in the Galactic plane.
\mgro\/ spatially associates with an IceCube hotspot of neutrino emission.
Although the hotspot is not significant yet, this suggests a possible hadronic origin of the observed gamma-ray radiation.
Here we describe a multiwavelength analysis on \mgro\/ to determine its nature.
We identify, for the first time, an extended GeV source as the counterpart of MGRO J1908+06, discovering possibly associated molecular clouds (MCs).
The GeV spectrum shows two well-differentiated components: a soft spectral component below $\sim10$ GeV, and a hard one ($\Gamma\sim1.6$) above these energies.
The lower-energy part is likely associated with the dense MCs surrounding the supernova remnant \snr\/, whereas the higher-energy component, which connects smoothly with the spectrum observed in TeV range, resembles the inverse Compton emission observed in relic pulsar wind nebulae.
This simple scenario seems to describe the data satisfactorily, but raises questions about the interpretation of the emission at hundreds of TeV.
In this scenario, no detectable neutrino flux would be expected.

\end{abstract}

\keywords{Gamma-rays: pulsars: individual (PSR J1907+0602)}

\section{Introduction}
\label{intro}

\mgro\/ is an extended bright TeV source of unknown nature.
It was first discovered by the Milagro water Cherenkov telescope in its sky survey results after seven years of operation (Abdo et al. 2007).
\mgro\/ was subsequently detected in the TeV range by the High Energy Stereoscopic System (H.E.S.S., Aharonian et al. 2009), the Very Energetic Radiation Imaging Telescope Array System (VERITAS, Ward 2008; Aliu et al. 2014), the Astrophysical Radiation with Ground-based Observatory at YangBaJing (ARGO-YBJ, Bartoli et al. 2012), and recently by the High-Altitude Water Cherenkov Observatory (HAWC, Abeysekara et al. 2017).
The TeV luminosity of \mgro\ is comparable to the Crab Nebula (Bartoli et al. 2012), making it one of the most luminous Galactic gamma-ray sources in the TeV range.
MGRO J1908+06 is among the four sources detected above 100 TeV in a recent study by HAWC, indicating its possible ability to accelerate particles to PeV energy (Abeysekara et al. 2020).
Besides its possible PeVatron nature, \mgro\/ is spatially associated with an IceCube neutrino emission hotspot (Aartsen et al. 2019; Aartsen et al. 2020), though the post-trial significance is low.

The nature of  \mgro\/ remains unrevealed.
Searches for its multiwavelength counterparts in radio, X-ray and GeV gamma rays have been unsuccessful (Abdo et al. 2010; Kong et al. 2007; Pandel 2015; Duvidovich et al. 2020).
\mgro\/ is spatially associated with a middle-aged (20-40 kyr, Downes et al. 1980) supernova remnant (SNR) G40.5$-$0.5 and an energetic gamma-ray pulsar \psrj\/, with a spin period of 107\,ms, a spin-down luminosity of $\sim 2.8 \times 10^{36}\,$erg s$^{-1}$ and a characteristic age of 19.5 kyr.
The distance to \psrj\/, estimated from its dispersion measure, is 3.2$\pm$0.6 kpc (Abdo et al. 2010a), and distance estimates of \snr\/ place it either in a similar neighborhood (3.5 kpc, using CO observations, Yang et al. 2006) or a more distant region (using the $\Sigma$-D relation, 5.5 to 8.5 kpc, Downes et al. 1980; 6.1 kpc, Case \& Bhattacharya 1998).
\psrj\/ was not likely born in \snr\/, considering the high estimated transverse velocity, and the lack of bow shock \& trail morphology.
The recent discovery of the radio pulsar \psrjj\ (7.9 kpc, Lyne et al. 2017) in the projected center of the remnant promotes a larger distance estimation for \snr\/\footnote{using the electron-density model of Yao et al. (2017) (\url{http://119.78.162.254/dmodel/index.php}), the distances of \psrj\/ and \psrjj\/ estimated from dispersion measure are nearer, being ~2.6 kpc and ~6.7 kpc, respectively.}.
Besides \psrj\/, GeV source 4FGL J1906.2+0631 is also spatially associated with MGRO J1908+06 (Abdollahi et al. 2020).
Below we summarize these multiwavelength spatial associations.
\begin{itemize}
  \item \psrj\/, GeV and radio pulsar, distance $\sim$ 3.2 kpc
  \item 4FGL J1906.2+0631, unidentified GeV source, distance unknown
  \item \snr\/, radio supernova  remnant, likely distance $\sim$ 8 kpc
   \item \psrjj\/, radio pulsar, likely associated with \snr\/, distance $\sim$ 8 kpc

\end{itemize}

The nature of MGRO J1908+06 is under debate.
A leptonic pulsar wind nebula (PWN) scenario powered by PSR J1907+0602  and a hadronic scenario powered by SNR G40.5-0.5 have been proposed (Abdo et al. 2010a).
Duvidovich et al. (2020) proposed that the TeV emission of \mgro\/  originates from a combination of the two scenarios above which play their own roles at different distances.
To gain further insight into these scenarios, we have analyzed multiwavelength observations in the vicinity of \mgro\/ and modelled the emission processes, and we report our results below.

\section{Multiwavelength observations}

Multiwavelength observations were investigated in this paper, including CO observations from the Milky Way Imaging Scroll Painting (MWISP) project (Su et al. 2019), \emph{Fermi} Large Area Telescope (LAT) observations of GeV gamma rays, \xmm\/ observations in X-ray, and the Very Large Array Galactic Plane Survey (VGPS, Stil et al. 2006) in radio.
The details of these observations are reported in Appendices.

\section{Results}

\subsection{Molecular clouds in the region of \mgro\/}

Molecular clouds (MCs) towards the \snr\ region have been studied in the past (Yang et al. 2006; Duvidovich et al. 2020).
Using data from the MWISP project (Su et al. 2019), we searched for MCs with \twCO\ (\otz),\thCO (\otz), and C$^{18}$O ($J$=1-0)
emission lines in a larger region covering \mgro\/.
MCs are discovered to be spatially associated with \mgro\/ in the \twCO\ (\otz) and \thCO\ (\otz) maps
between 46-66 km/s (Figure \ref{co}; also see \emph{Appendix A} for details), which is consistent with the possible distance, $3.2$ kpc, to \snr\ and \psrj\/.
A shell-like cavity encircling the radio morphology of \snr\/ is observed in both \twCO\ (\otz) and \thCO\ (\otz) maps, indicating a possible SNR swept-up shell (Figure \ref{co}).

We show the \twCO\ (\otz) and \thCO\ (\otz) maps for five consecutive velocity ranges from 46$\km\ps$ to 66$\km\ps$ with a coverage of 4$\km\ps$ in \emph{Appendix A}, Figure \ref{COvelosities}.
Different regions with MCs are studied in detail and their astrophysical properties are estimated in \emph{Appendix A}.
The mean density is estimated to be $\sim$ 45 ${\rm cm}^{-3}$.
In the 46$\km\ps$ to 66$\km\ps$ velocity slices, there are apparent filaments of MCs positionally located next to \psrj\/.
We have inspected the $^{12}$CO (J=1$-$0) and $^{13}$CO (J=1$-$0) line profiles of the MCs toward the \snr\/ and \psrj\/ region, to search for kinematic  evidence for gas distribution due to external interaction (e.g., Zhou \& Chen, 2011; Liu et al., 2020).
However, we do not find any such evidence of asymmetric broad profiles of the $^{12}$CO line  (i.e., a wing part deviating from the main Gaussian in a $^{12}$CO profile where the signal-noise-ratio of the $^{13}$CO is less than 3) in the grid of the CO spectra.
For a further search of interaction signals, we studied the data from the INT Photometric H$\alpha$ Survey of the Northern Galactic Plane (IPHAS) covering the \mgro\/ region.
No H$\alpha$ ascribable to interaction was detected.
%

\begin{center}
\begin{figure*}
\centering
\includegraphics[scale=0.42]{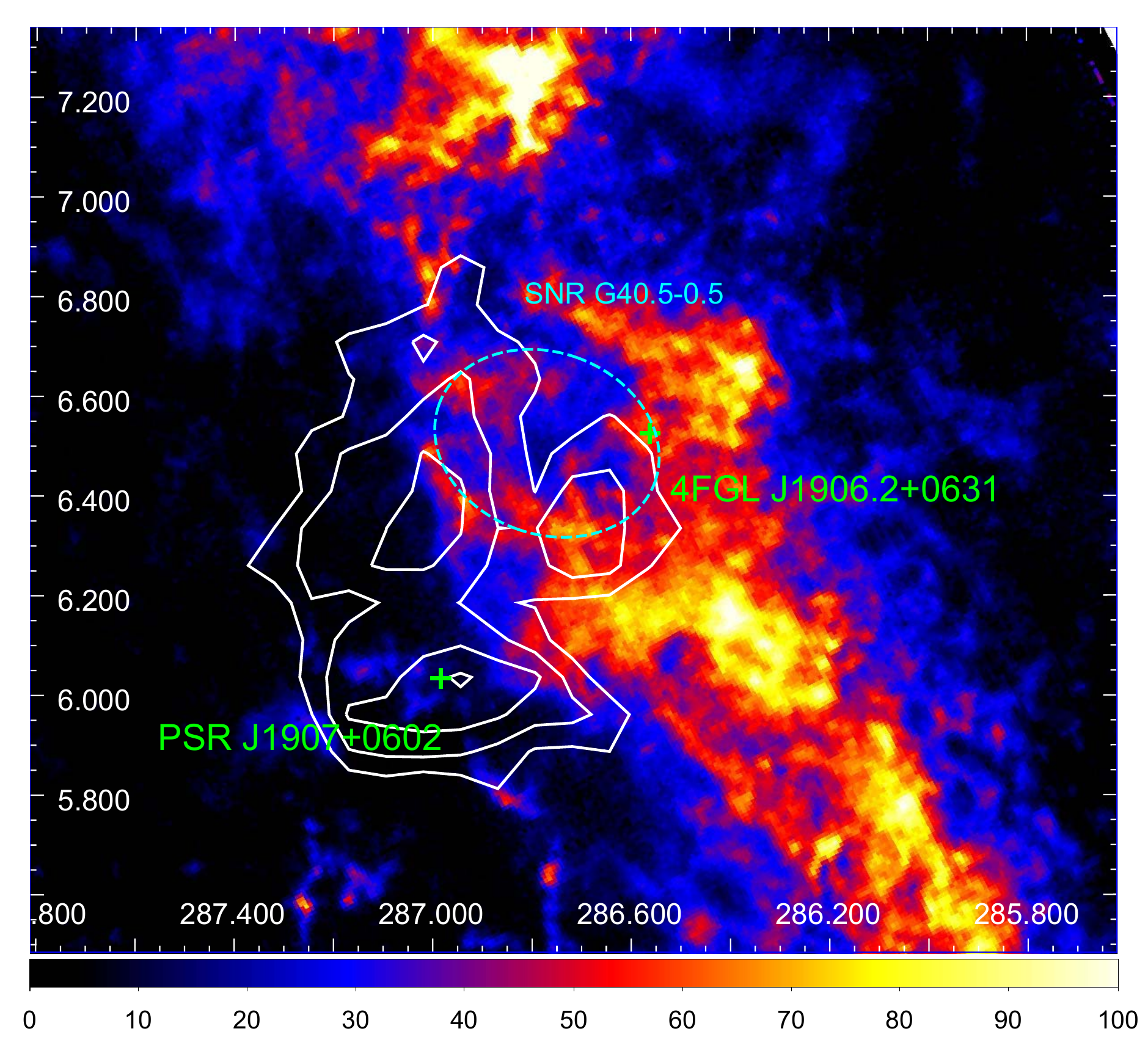}
\includegraphics[scale=0.42]{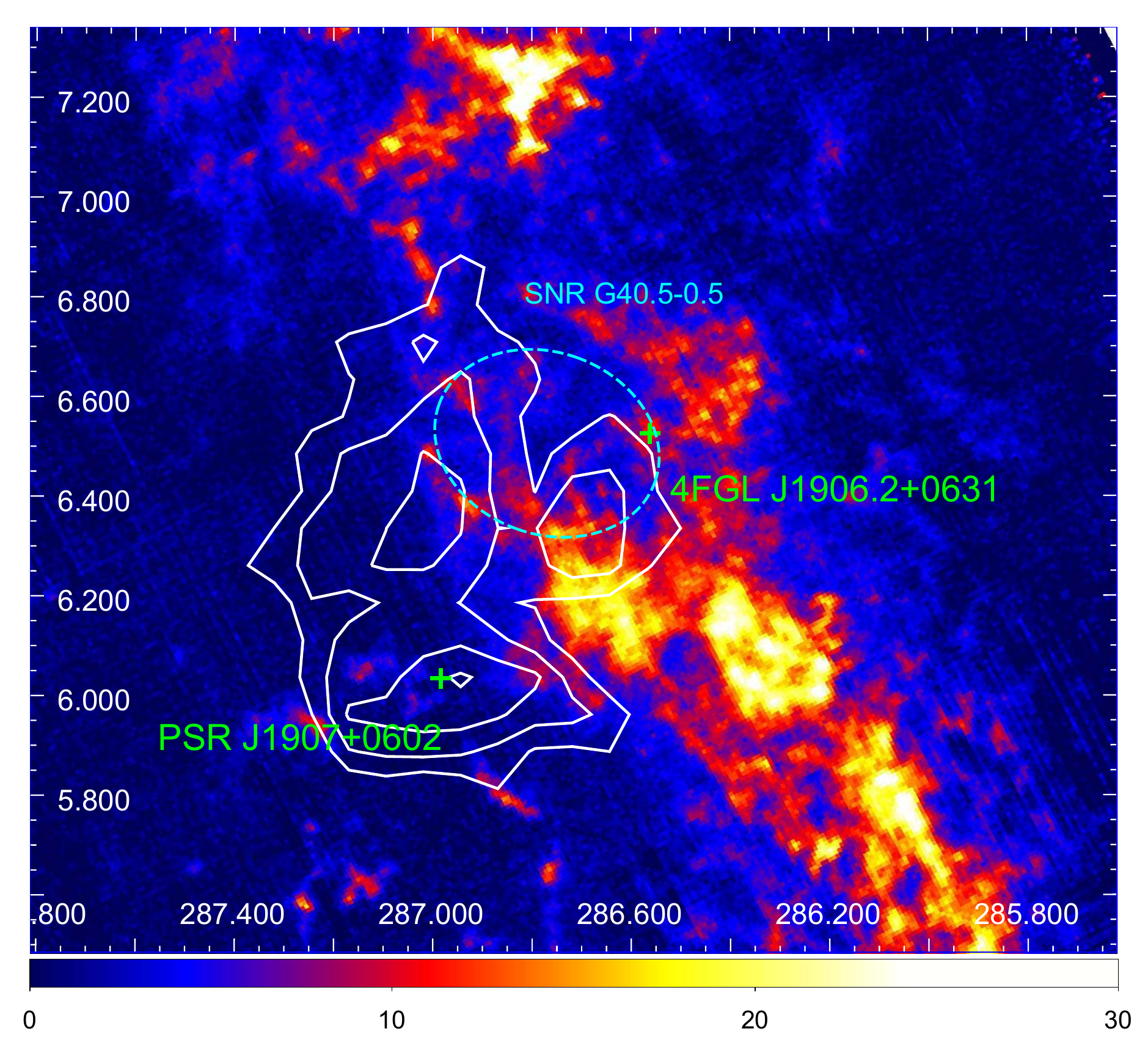}

\caption{\twCO\ (\otz) (left) and \thCO\ (\otz) (right) intensity maps (in the unit of K $\km\ps$) integrated in the velocity range $46$--$66\km\ps$.
4FGL J1906.2+0631 and \psrj\ from 4FGL are shown with green crosses while the radio morphology of \snr\/ is indicated with a dotted ellipse.
White contours correspond to the VERITAS significance map (Aliu et al. 2014) ranging from 3$\sigma$ to 6$\sigma$ by 1$\sigma$ steps.
The x and y axes are R.A. and decl. (J2000) in degrees.
}
\label{co}
\end{figure*}
\end{center}

\subsection{GeV counterpart of \mgro\/}

\begin{figure}
\centering
\includegraphics[scale=0.4]{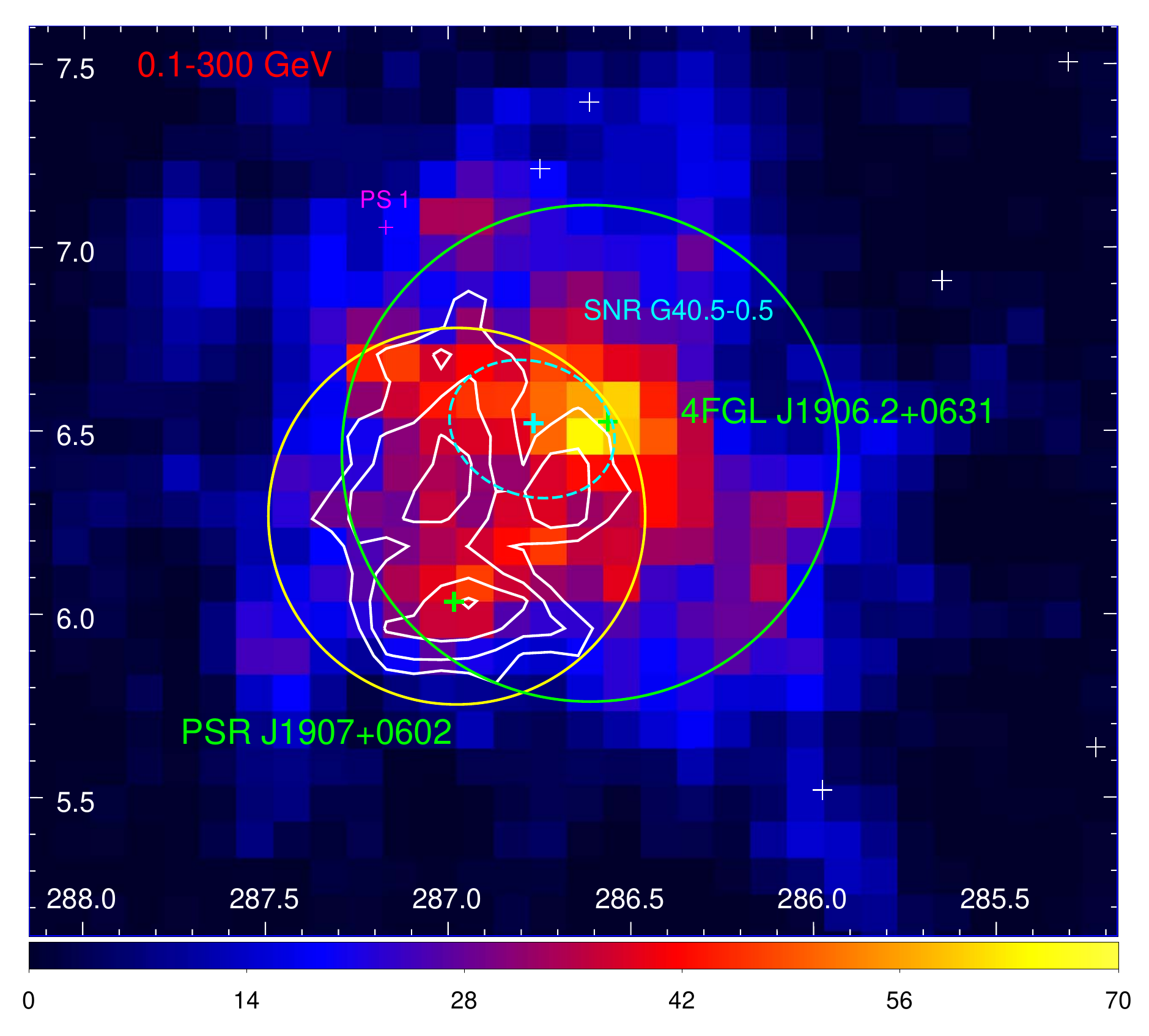}
\includegraphics[scale=0.39]{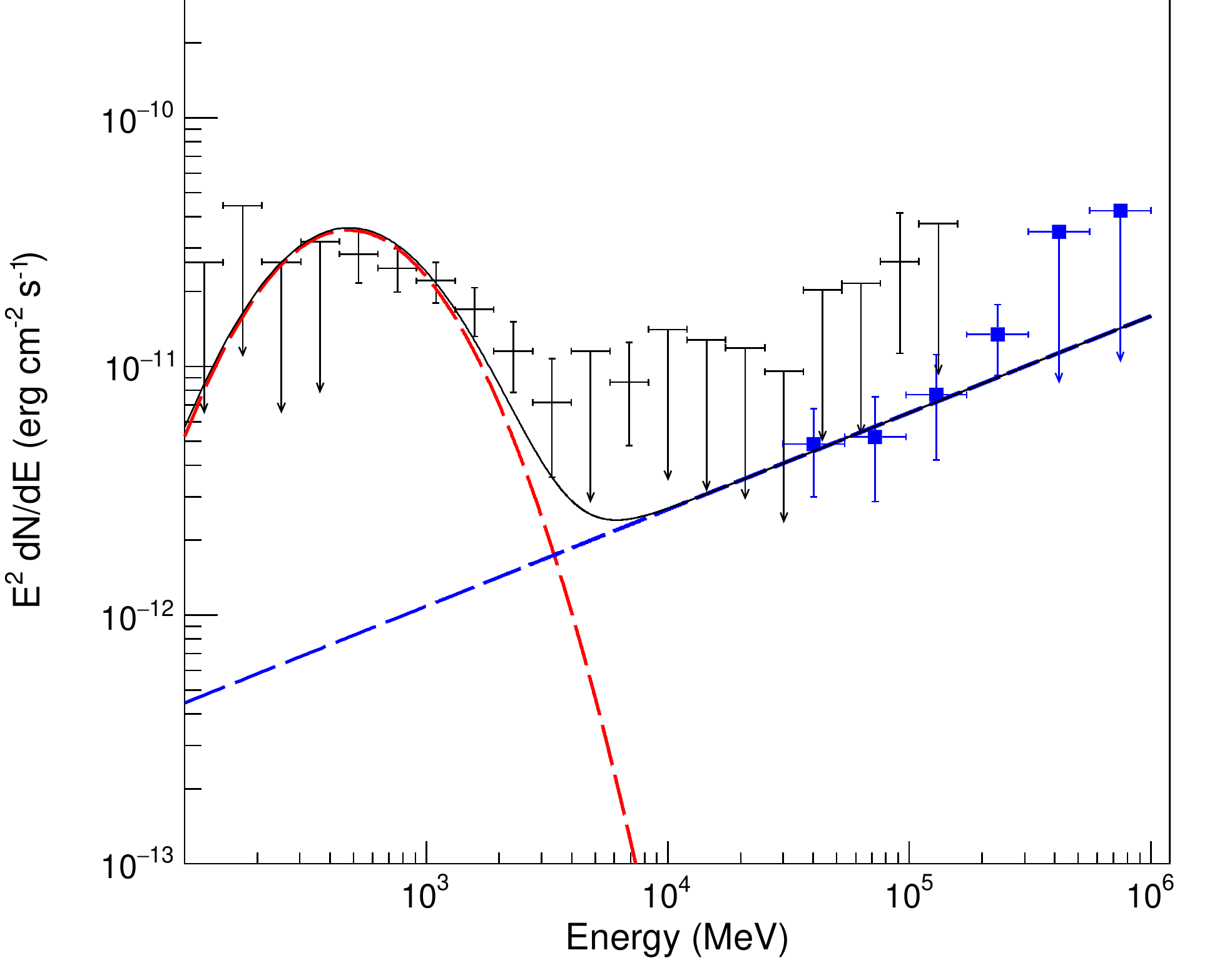}\\
\centering
\includegraphics[scale=0.4]{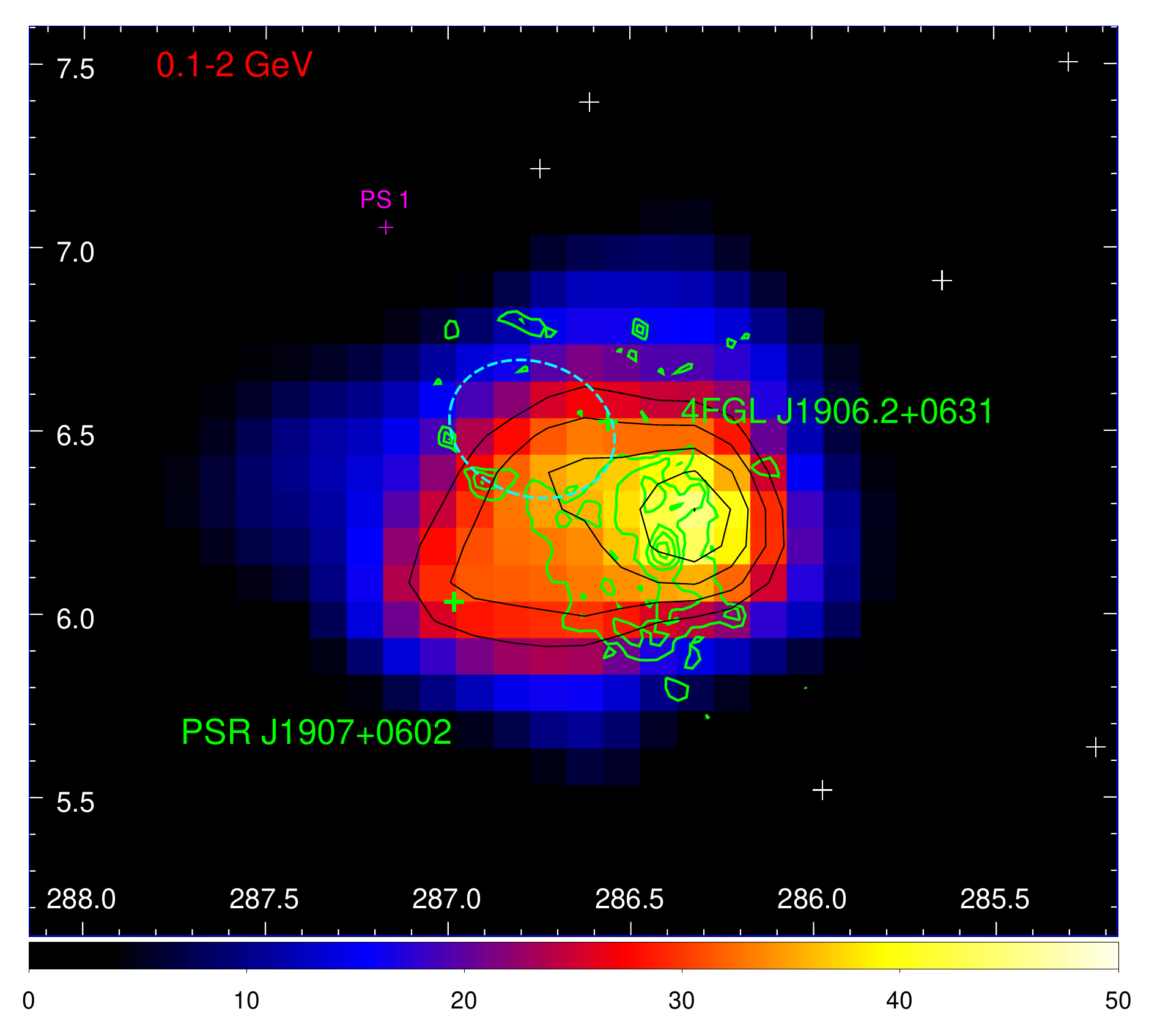}
\includegraphics[scale=0.4]{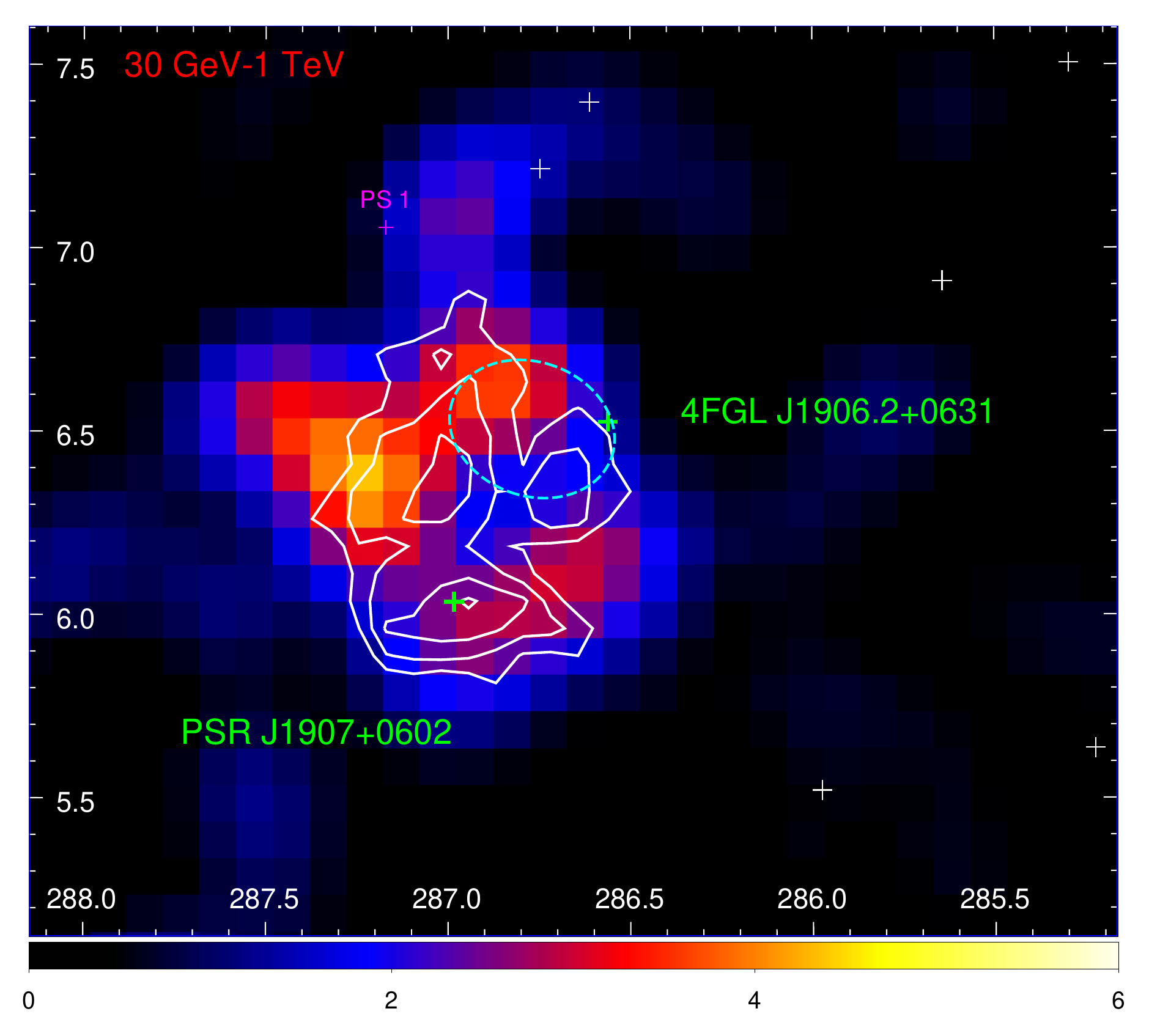}

\caption{\textbf{Top left}: \emph{Fermi}-LAT TS map of \mgro\/ region in the 0.1--300 GeV range.
Background 4FGL sources are shown with white crosses.
For comparison, we show the 68\% containment radius of disk morphology determined in 1--300 GeV and Gaussian morphology measured by H.E.S.S. (Aharonian et al. 2009) with green and yellow circles, respectively (r$_{68, disk}$=0.82r$_{disk}$=0.68\degree; r$_{68, Gaussian}$=1.51$\sigma_{Gaussian}$=0.51\degree; Lande et al. 2012).
The magenta cross indicates the new source PS 1.
\psrj\ and 4FGL J1906.2+0631 are shown with green crosses.
\snr\/ is indicated with a dotted cyan ellipse and the possibly associated pulsar \psrjj\ is shown with a cyan cross.
Other labels are as in Figure \ref{co}.
The x and y axes are R.A. and decl. (J2000) in degrees.
\textbf{Top right}: GeV SED of \nsource\/.
Data points in black and blue are from the \psrj\/ off-peak and phase averaged analyses, respectively (see text for detail).
The dashed red and blue curves on two instances indicate the two-component spectral modelling (Table \ref{fit}) with the sum shown with a black line.
\textbf{Bottom left}: \emph{Fermi}-LAT TS map of \mgro\/ region in the 0.1--2 GeV range.
The black contours correspond to TS values staring from 25 with a step size of 5.
The green contours correspond to the \twCO\ (\otz) 62 -- $66\km\ps$ intensity map of the surrounding region, starting from 15K with a step of 5K.
The labels are as in the top left panel.
\textbf{Bottom right}: \emph{Fermi}-LAT TS map of \mgro\/ region from 30 GeV--1 TeV smoothed with 0.15 degree Gaussian.
The labels are as in the top left panel.}
\label{tsmap}
\end{figure}

Though being very bright at TeV energies, no GeV counterpart for \mgro\/ was identified in previous \emph{Fermi}-LAT studies (Abdo et al. 2010a; Ackermann et al. 2011).
Two gamma-ray sources \psrj\/ and \fgl\/ are spatially associated with \mgro\/, as listed in the 4FGL catalog (Abdollahi et al. 2020).
For a deeper search of its GeV counterpart, we analyzed more than 11 years of \emph{Fermi}-LAT data in the 0.1--300 GeV band.
To minimize the contamination from the bright gamma-ray pulsar PSR J1907+0602, we carried out data analysis during the off-peak phases of this pulsar.
A full description of our data analysis is presented in \emph{Appendix B}.
We discovered a previously-undetected, extended GeV source spatially associated with \mgro\/, hereafter referred as \nsource\/, which is shown in Figure \ref{tsmap} together with the TeV extension measured by H.E.S.S. and the TeV contours reported by VERITAS.
The good morphological agreement between \nsource\ and \mgro\ strongly suggests a common origin. To investigate this, we derived the morphological and spectral properties of \nsource\/.

A morphology study was carried out in 1-300 GeV to estimate the extension of \nsource\/ assuming a power-law spectral model.
We tested both the disk and Gaussian morphologies using \textit{Fermipy}, and the disk morphology yields more significant extension detection.
The disk morphology resulted in a center position of R.A.=286.61\degree $\pm$0.08\degree, Decl.= 6.44\degree $\pm$0.07\degree, and a radius of 0.83\degree $\pm$0.05\degree, yielding a TS$_{ext}$=44 (see \emph{Appendix B}).
Comparing the likelihood values, the disk morphology is significantly preferred over a two-point-source model (\psrj\/ plus \fgl\/ from 4FGL) with a $\Delta$TS=23 (see Appendix B).
We further considered a two-component spatial model, constituted by a template using VERITAS counts map (Aliu et al. 2014) plus the point source \fgl\/.
However, comparing the likelihood values, the disk morphology is again significantly preferred over this two-component spatial model with a $\Delta$TS=18.
Additionally, in this two-component spatial model \fgl\/ is not significantly detected either in 1-300 GeV or 0.1-300 GeV, leading to a TS value of 11 and 20, respectively, which are lower that the detection threshold (TS=25, 4FGL).
Very recently, Di Mauro et al  (2020) carried out an analysis of \mgro\/ and detected extended emission, which is consistent with the results shown here.

We carried out the spectral analysis in 0.1-300 GeV.
Adopting the disk morphology, we tested a power-law ($dN/dE=N_{0}(E/E_{0})^{-\Gamma}$ cm$^{-2}$ s$^{-1}$ MeV$^{-1}$) and a log-parabola spectral model ($dN/dE=N_{0}(E/E_{b})^{-(\alpha+\beta log(E/E_{b}))}$ cm$^{-2}$ s$^{-1}$ MeV$^{-1}$).
The log-parabola spectral model is preferred with a $\Delta$TS=89, indicating a significant spectral curvature.
Low-energy turnovers are expected for proton-proton interactions, indicating a hadronic nature for \mgro\/.
With the disk morphology and log-parabola spectral model, \nsource\ is significantly detected with a TS value of 190 and an energy flux of (7.44$\pm$0.64) $\times$ 10$^{-11}$ erg~cm$^{-2}$s$^{-1}$ between 0.1$-$300 GeV\footnote{To explore the influence of different Galactic diffuse emission models, we tested the previous Galactic diffuse emission component (``gll\_iem\_v06.fits"; Aecro et al. 2015) in our background spectral-spatial model. \nsource\ is again significantly detected but with a lower TS value of 155 and a consistent energy flux of (7.30$\pm$0.34) $\times$ 10$^{-11}$ erg~cm$^{-2}$s$^{-1}$ between 0.1$-$300 GeV.}.
Adopting the disk morphology, the best-fitted parameters are shown in Table \ref{fit} and the gamma-ray spectral energy distribution (SED) of \nsource\ is shown in Figure \ref{tsmap}, top right panel (in black).
An excess beyond $\sim$5 GeV, deviating from the log-parabola spectrum, suggests the existence of another spectral component which may originate from a second population of accelerated particles, or from a second emitting region.

\begin{center}
\begin{figure*}
\centering
\vspace{-2.5cm}
\includegraphics[scale=0.8]{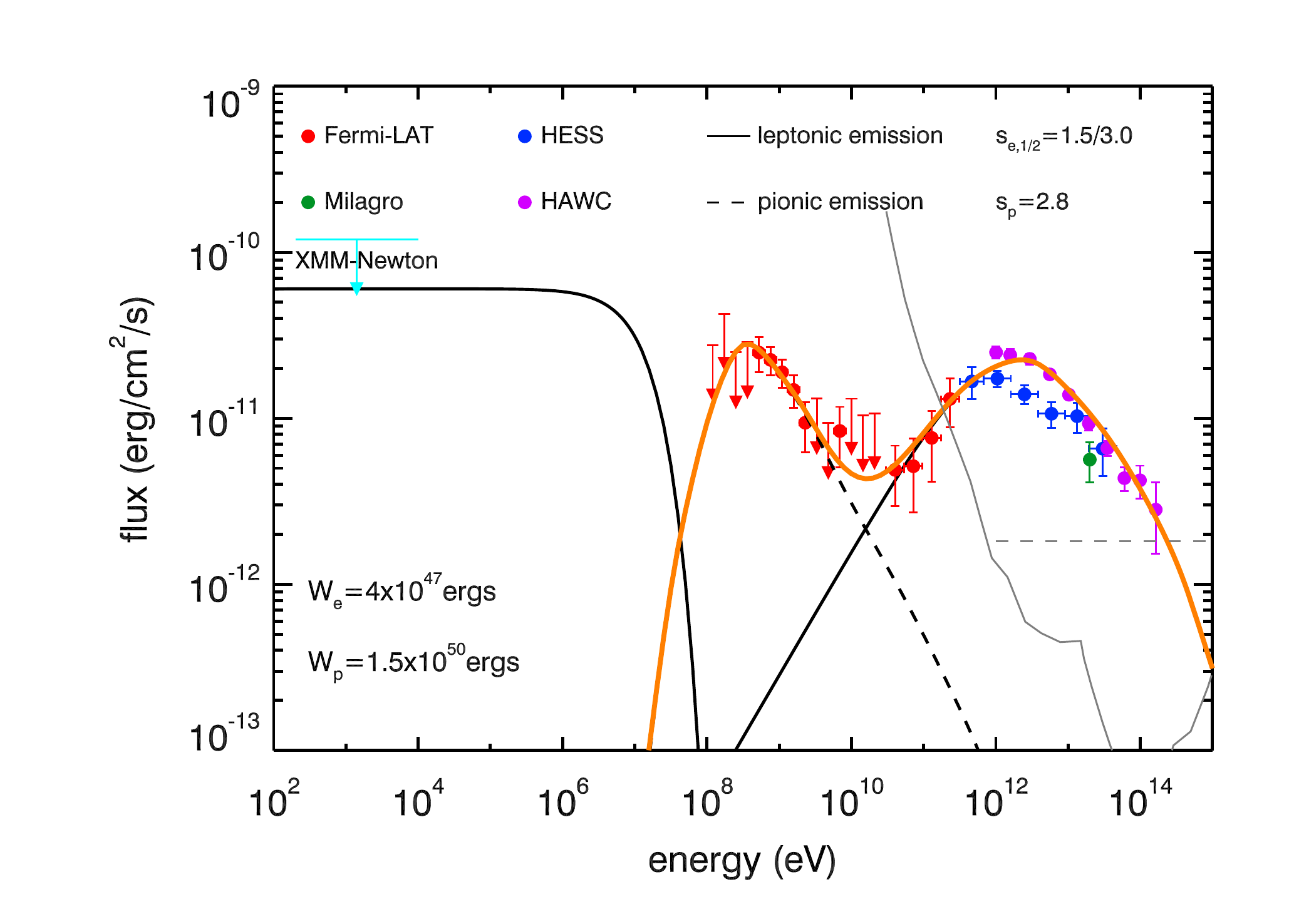}
\caption{Multi-wavelength SED of \mgro\/ with hadronic and leptonic hybrid modeling.
Besides the GeV and X-ray measurements in this paper, other data are taken from Abdo et al. (2007) (Milagro), Aharonian et al. (2009) (H.E.S.S.), and Abeysekara et al. (2020) (HAWC).
The VERITAS SED data are consistent with H.E.S.S. and are not shown for concision.
 The solid grey curve shows the LHAASO point-source sensitivity of one-year exposure (Bai et al. 2019).
 The dashed grey curve represents the upper limit for pionic gamma-ray emission accompanying the neutrino emission (Aarsten et al. 2020).
Please see Section 4 for details.
}
\label{multiSED}
\end{figure*}
\end{center}

For higher statistics, we studied \emph{Fermi}-LAT data above 30 GeV further without pulsar gating.
With a spectral cutoff at 2.9 GeV, the magnetospheric emission from \psrj\/ is negligible above 30 GeV (Abdo et al. 2013), making pulsar gating unnecessary.
In the 30 GeV to 1 TeV energy range \nsource\ is detected as an extended source (Figure \ref{tsmap}, bottom right panel).
The best fitted disk model is centered at R.A.=286.88\degree $\pm$0.05\degree, Decl.= 6.29\degree $\pm$0.05\degree with a radius of 0.51\degree $\pm$0.02\degree, yielding a TS of 47 assuming a power-law spectral model and a TS$_{ext}$=46.
Considering the possible spectral connection with the TeV range, we tested the template produced from VERITAS counts map (Aliu et al. 2014).
Assuming a power-law model, \nsource\ is detected from 30 GeV--1 TeV with a TS of 43, which is a worse fit than a disk model.
Adopting the disk model, \nsource\ shows a spectral index of 1.61$\pm$0.16 and an energy flux of (3.05$\pm$0.70) $\times$ 10$^{-11}$ erg~cm$^{-2}$s$^{-1}$ (Table \ref{fit}), confirming the existence of an additional spectral component at higher energies beyond the log-parabola component (Figure \ref{tsmap}, top right panel).
We added this component to the analysis  of \nsource\ from 0.1--300 GeV in off-peak data, fixed it and constructed a two-component spectral model (log-parabola plus power law).
The results are reported in Table \ref{fit} and shown in Figure \ref{tsmap}, top right panel.
The multiwavelength SED of \mgro\/ is shown in Figure \ref{multiSED}.
The GeV SED shows a natural continuity to the TeV range.
The TS map in the 30 GeV to 1 TeV energy range is shown in Figure \ref{tsmap}, bottom right panel.
No clear \twCO\ (\otz) emission is associated with gamma-ray emission.

Below 2 GeV, the SED of \nsource\ is dominated by the log-parabola component.
Assuming a disk morphology (Figure \ref{tsmap}, bottom left panel), the best fitted values are R.A.=286.71\degree $\pm$0.07, Decl.= 6.30\degree $\pm$0.07 with a radius of 0.68\degree $\pm$0.07, yielding a TS value of 109.
The peak of gamma-ray emission is spatially consistent with the \twCO\ (\otz) intensity in the 62 -- $66\km\ps$ range, suggesting a hadronic origin.
We produced a template from the 62 -- $66\km\ps$ \twCO\ (\otz) intensity map within the best-fitted disk morphology.
The template leads to a better fitting with TS value of 113, which is preferred over the disk morphology but not by a large margin.

\begin{table*}{}
\centering
\scriptsize
\caption{Best-fit Spectra Parameters of \nsource\/. }
\begin{tabular}{lccccc}
\\
\\
 & &   Off-peak Analysis from 0.1-300 GeV  &  &  &  \\
   \\
\hline\hline
          Model                 &       $\Gamma$                          &   $\alpha$                         &     $\beta$                             &  Energy Flux                                                                    &  TS     \\
                                       &      (Power law)              &    (Log-parabola, E$_{b}$=1 GeV)                 &     (Log-parabola, E$_{b}$=1 GeV)                   & (10$^{-11}$ erg~cm$^{-2}$s$^{-1}$)                 &            \\
\hline              

Log-parabola                                        &  --                               &       2.90  $\pm$ 0.20         &      0.65 $\pm$ 0.11         &         7.44 $\pm$ 0.64          & 190    \\
(0.83\degree disk)                                                      &                                  &                                              &                                             &                                                &            \\

\hline\                 
Log-parabola $+$ Power law (fixed)                       & 1.61 (fixed)               &    3.16 $\pm$ 0.24           &       0.78 $\pm$ 0.13           &         9.34 $\pm$ 0.64       &  197   \\
(0.83\degree disk)  \qquad     (0.51\degree disk)      &                                   &                                              &                                             &                                                &            \\

\hline\hline\
\\
\\
 & &   Analysis from 30 GeV--1 TeV  &  &  &  \\
\hline\hline
          Model                 &         $\Gamma$                           &   $\alpha$                         &     $\beta$                             &  Energy Flux                                                                    &  TS     \\
                                       &      (Power law)              &    (Log-parabola, E$_{b}$=1 GeV)                 &     (Log-parabola, E$_{b}$=1 GeV)                   & (10$^{-11}$ erg~cm$^{-2}$s$^{-1}$)                 &            \\

\hline              
Power law                   & 1.61 $\pm$ 0.16              &   --        &      --          &         3.05 $\pm$ 0.70         &  47    \\
(0.51\degree disk)      &                                   &                                              &                                             &                                                &            \\

\hline\hline\
\\
\\

\label{fit}
\end{tabular}
\end{table*}

\subsection{Search for \mgro\/ counterparts in X-ray and radio wavelengths}

\begin{center}
\begin{figure*}
\centering
\includegraphics[scale=0.4]{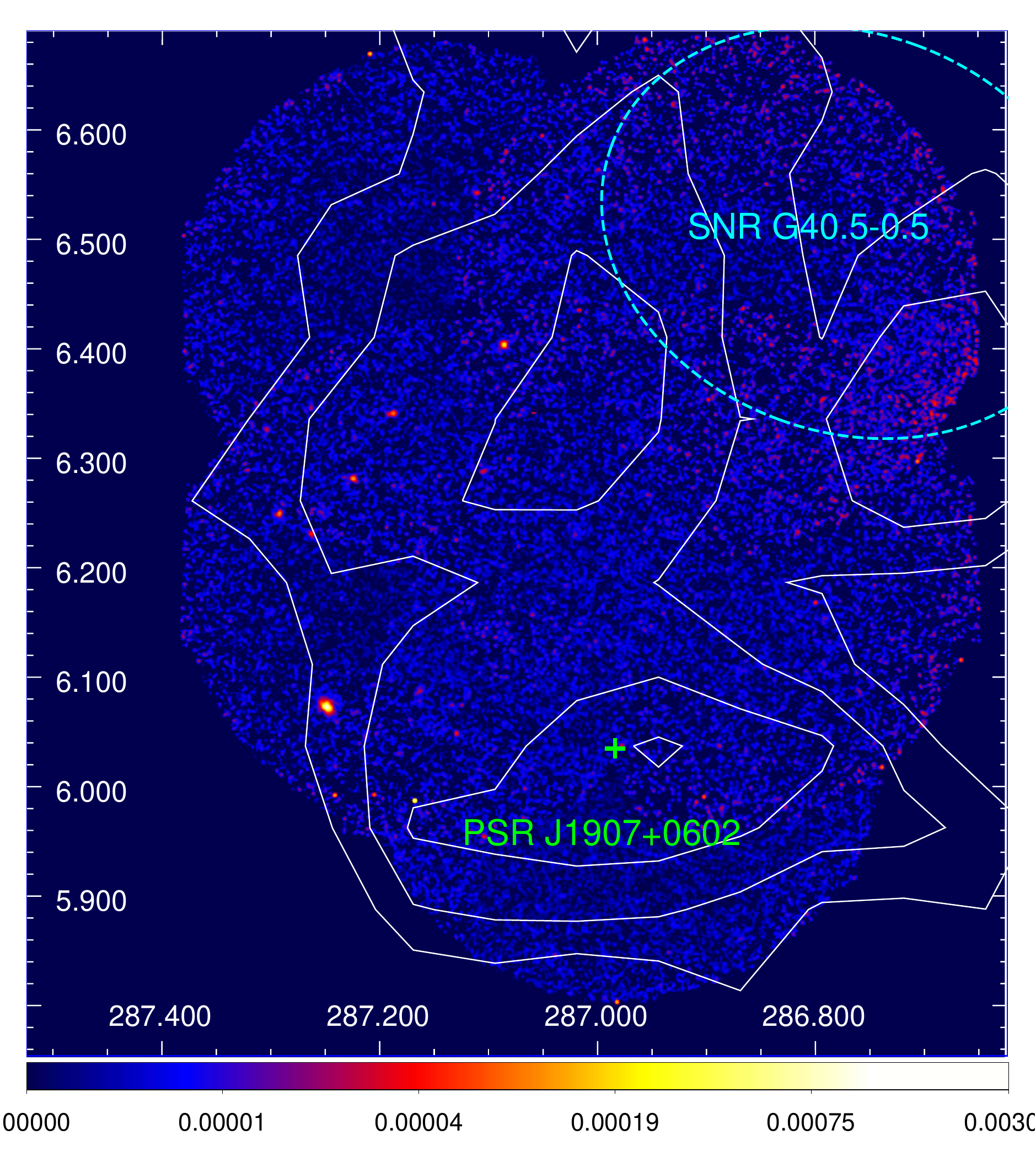}
\includegraphics[scale=0.4]{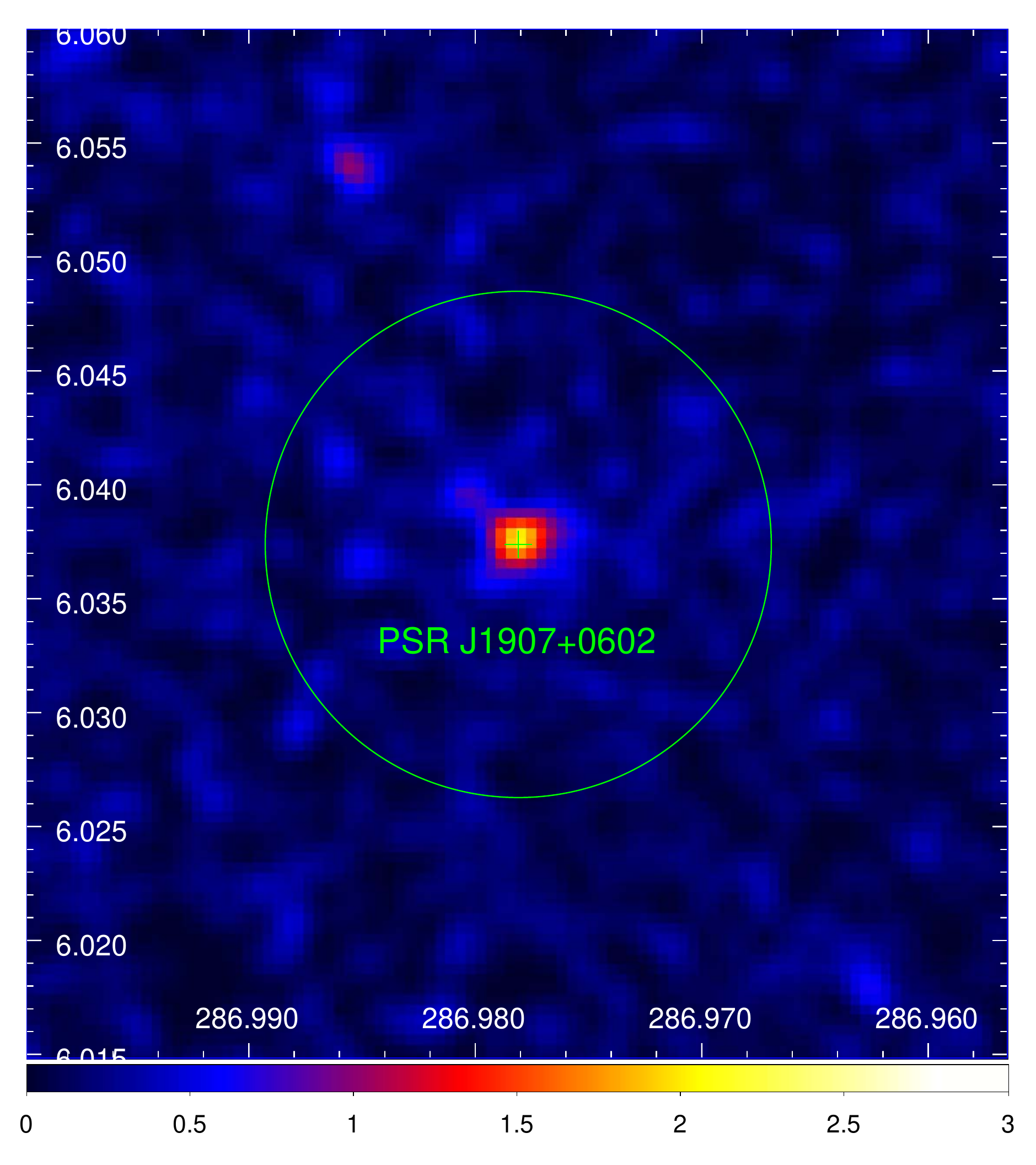}
\includegraphics[scale=0.4 ]{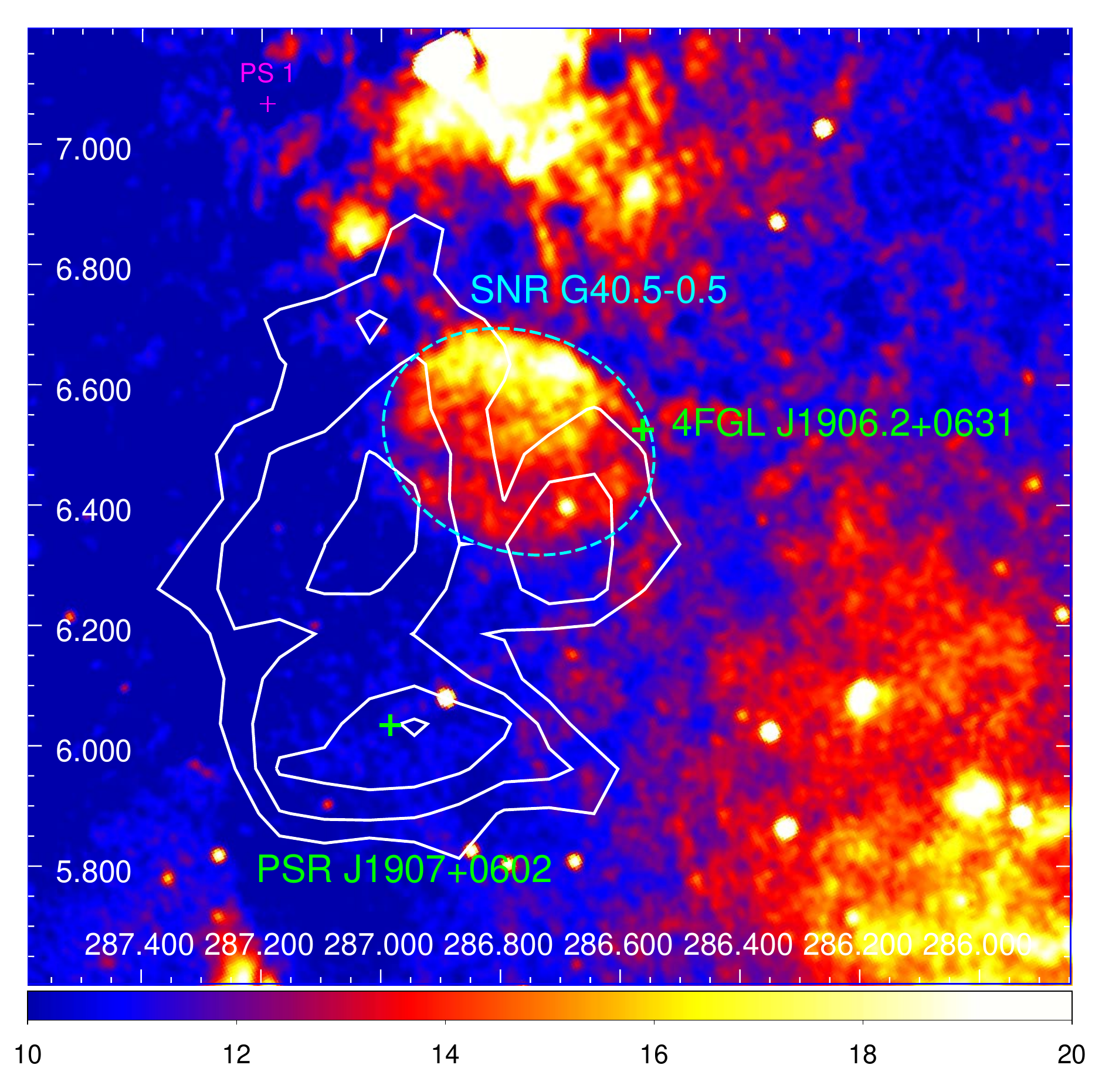}

\caption{\textbf{Top left}: Gaussian-smoothed ($\sigma$=9 arcsec) log-scaled, particle background-subtracted, and exposure-corrected count rate map of the \mgro\/ region from 0.2$-$10 keV, combining all available \xmm\ MOS 1 \& 2 data.
White contours are the VERITAS significance map, as in Figures \ref{co} and \ref{tsmap}.
The labels are as in Figure \ref{tsmap}.
\textbf{Top Right}: Gaussian-smoothed ($\sigma$=4.8 arcsec) \xmm\ MOS 1 \& 2 combined counts map of \psrj\ region from 2$-$8 keV.
The green cross shows the pulsar timing position (Abdo et al .2010a).
 The green circle with a radius of 40 arcsec indicates the radius of $\sim$ 90\% fractional encircled energy of MOS 1 and MOS 2, which should contain $\sim$ 90\% of the counts from a point source.
 \textbf{Bottom}: VGPS 1420 MHz image of the \mgro\/ region.
Legends are the same as the first panel.
 }
\label{XMM}
\end{figure*}
\end{center}

The \XMM\/ X-ray satellite has covered \mgro\/ with 5 observations (Obs ID 0553640101, 0553640201, 0553640701, 0553640801 and 0605700201), providing a total exposure of 109 ks.
Combining all available MOS data, we produced a particle background-subtracted and exposure-corrected count rate map for the \mgro\/ region (Figure \ref{XMM}, left panel).
No diffuse X-ray emission coincident with the gamma-ray emission region of \mgro\/ could be detected in the 0.2 to 10 keV energy range.
The morphology measured by H.E.S.S. (a Gaussian profile with $\sigma$ of 0.34$\degree$; Aharonian et al. 2009) is the only one in TeV range whose 1$\sigma$ extension is fully covered by these \XMM\/ observations.
 Considering the 1$\sigma$ range of H.E.S.S. morphology, excluding point sources within, assuming a spectral index of 2 and a H\,{\sc i} column density of N$_{H}$=1.3 $\times$10$^{22}$cm$^{-2}$ (Abdo et al. 2010a), we calculated a 95\% unabsorbed upper limit for \mgro\/ as 1.2 $\times$ 10$^{-10}$ erg~cm$^{-2}$s$^{-1}$ (0.2$-$10 keV).
From our CO observation, the local H\,{\sc i} column density to the \mgro\/ region is N$_{H}$= 2$\times$N$_{H_{2}}\sim(2-7)\times10^{21}$cm$^{-2}$ (\emph{Appendix A}, Table \ref{COparameter}), which is much lower than the  total Galactic H\,{\sc i} column density in this direction estimated using the HEASARC tool\footnote{\url{https://heasarc.gsfc.nasa.gov/cgi-bin/Tools/w3nh/w3nh.pl}} and with \CXO\/ (N$_{H}$=1.3 $\times$10$^{22}$cm$^{-2}$; Abdo et al. 2010a), suggesting that the local absorption does not likely lead to the lack of an X-ray counterpart.

With a 19 ks \CXO\ observation (obs. ID 7049 on 2009-08-19), Abdo et al. (2010a) reported the detection of \psrj\/ and hints of extension from 2$-$8 keV.
However, because of the low statistics, no PWN associated with \psrj\/ could be unambiguously identified.
In the \xmm\/ data, Obs. ID 0605700201 (on 2010-04-26, 52.6 ks exposure), ID 0553640201 (on 2008-10-02, 24.7 ks exposure), ID 0553640701 (on 2009-03-19, 7.4 ks exposure)  have covered \psrj\/.
However, \psrj\ is only detected in obs. ID 0605700201, which provided the longest exposure (52.5 ks), as a point source, by combining MOS 1 and MOS 2 data.
To check the previously claimed possible extension by \emph{Chandra} (Abdo et al. 2010a), we produced counts map from 2$-$8 keV of \psrj\/, combining MOS 1 and MOS 2 data.
All counts are well located within the radius of $\sim$ 90\% fractional encircled energy\footnote{\url{https://xmm-tools.cosmos.esa.int/external/xmm_user_support/documentation/uhb/}} (Figure \ref{XMM}, top right).
The detected emission is thus consistent with a point source and no associated PWN could be identified.

We checked for variability with spectral analysis of the eight-month separated \xmm\/ (Obs. ID 0605700201 on 2010-04-26)  and \CXO\  (ObsID 7049 on 2009-08-19) observations.
We adopted a simple power-law model plus absorption with H\,{\sc i} column density fixed at N$_{H}$= 1.3$\times$10$^{22}$cm$^{-2}$ (Abdo et al. 2010a) because of the low overall counts.
The 1$-$10 keV \xmm\/ MOS 1\&2 data yield a best-fit spectral index of 1.79$^{+0.42}_{-0.41}$ with unabsorbed total energy flux of 6.29$^{+1.44}_{-1.18}$ $\times$10$^{-14}$ erg~cm$^{-2}$s$^{-1}$, whereas the \CXO\ ACIS spectrum is well fitted with compatible spectral parameters, a best-fit spectral index of 1.28$^{+0.45}_{-0.42}$ and unabsorbed total energy flux of 5.14$^{+1.47}_{-0.94}$ $\times$10$^{-14}$ erg~cm$^{-2}$s$^{-1}$.
No spectral or flux variability can be claimed.
The \xmm\/ and \CXO\/ spectra are shown in \emph{Appendix C}, Figure \ref{Xray}.

During this \xmm\ observation, the PN camera was operating in small window mode, providing sufficient time resolution (5.7 ms) to search for X-ray pulsations.
We extract photons from PN data using a radius of 20 arcsec in 0.2$-$10 keV and 3$-$10 keV with barycenter correction applied.
Using \emph{Tempo2} (Hobbs et al. 2006) and the \emph{photons} plugin\footnote{\url{http://www.physics.mcgill.ca/~aarchiba/photons_plug.html}} and contemporaneous \emph{Fermi}-LAT gamma-ray ephemeris, we have assigned pulsar rotational phase to each extracted photon.
No significant X-ray pulsation is detected.

The VGPS at 1420 MHz (Stil et al. 2006) has covered \mgro\/ region and the image is shown in Figure \ref{XMM}, bottom panel.
\psrj\/ and its possible associated PWN are not detected, but \snr\/ is clearly visible (Abdo et al. 2010a).
No diffuse radio large-scale emission associated with \mgro\/ is seen.
%

\section{Discussion}

Detailed analysis of the \emph{Fermi}-LAT data revealed that the gamma rays from the direction of \mgro\/ follow a two-component spectrum.
Based on the TS maps and multiwavelength observations, we propose two accelerators on the field: one region related to \snr\/, and a second one on the direction of the gamma-ray pulsar \psrj\/.
The high energy component discovered in \emph{Fermi}-LAT data shows a hard spectrum of $\alpha_{\rm he}\sim1.6$ and connects smoothly with the very-high energy spectrum measured by Cherenkov instruments (Figure \ref{multiSED}; Aliu et al. 2014; Aharonian et al. 2009; Abeysekara et al. 2020).
Previous studies (Abdo et al. 2010a; Abeysekara et al. 2017) have tentatively associated \mgro\ with a PWN powered by \psrj\ at a distance of 3.2\,kpc.
The high-energy ($>30$ GeV) component measured by \emph{Fermi}-LAT, combined with the very-high energy spectrum measured by Cherenkov instruments, do indeed resemble the spectral signature associated with inverse Compton emission from GeV/TeV PWNe (e.g. Crab, Abdo et al. 2010b; MSH 15-52, Abdo et al. 2010c; HESS J1825-137, Grondin et al. 2011).
The low-energy component (below a few GeV) is described by a soft spectrum, similar to the ones observed on evolved SNRs (Acero et al. 2016)\footnote{SNR analysis in Acero et al. (2016) starts from 1GeV. \nsource\ spectrum above 1 GeV could be well represented by a power-law with index of 2.69$\pm$0.19, which is consistent with SNRs' spectra reported in Acero et al. (2016)}, and it shows a significant peak coincident with an enhancement of molecular material (see Figure \ref{tsmap} Bottom Left and results shown in the \emph{Appendix}), implying a tentative hadronic origin.
Within 50 pc region of \snr\ and \psrjj\/, there locate several MCs which have compatible distances (Table \ref{COparameter}, far distance).
Considering that \psrjj\ is not a gamma-ray pulsar, we attempt to associate the emission to the CR-MC interaction for the low energy contribution at $\sim$8 kpc, whereas the high energy part of the spectrum is associated to leptonic emission associated to the putative PWN of \psrj\/ at $\sim$3.2 kpc.

To model the total emission in the region we assumed a combination of hadronic and leptonic emission, as is shown in Figure~\ref{multiSED}.
For the hadronic $pp$ collision, the gamma-ray flux is calculated using the parameterized formulae provided by Kamae et al. (2006), for a proton spectrum of the form of a power-law function in energy space, with a slope $s_p$.
Assuming a distance of 8\,kpc, the steep gamma-ray spectrum at low energy can be well-reproduced by employing $s_p=2.8$, which is consistent with the measured CR slope in the local interstellar medium.
Such a steep slope may reflect a diffusion-modifiled spectrum, i.e., the injection spectrum is softened by the energy-dependent diffusion of CR protons with a diffusion coefficient $D(E)\propto E^\delta$. We do not further specify the value of the injection spectral slope of CR protons and the index of the diffusion coefficient $\delta$, but just note that an injection spectral slope of $2.3-2.4$ and $\delta =0.4-0.5$ could be a reasonable combination of these two parameters, where the former one is motivated by the discovery of the pionic gamma-ray components from two other SNRs, i.e., W44 and IC~443 (Ackermann et al. 2013), and the latter one is based on the observation and modelling of secondary-to-primary CR ratios (Aguilar et al. 2016; Genolini et al. 2019, Huang et al. 2020).
We assume an average hydrogen density of $n=45\,\rm cm^{-3}$ in the surrounding MC, resulting in a $pp$ collision cooling timescale of $t_{pp}\simeq \left(0.5\sigma_{pp} nc\right)^{-1}=7\times 10^5(n/100\rm cm^{-3})^{-1}\,$yr. We multiply a factor of 1.4 to the obtained spectra of secondaries to account for the contribution of nuclei in molecular materials.
The total proton energy needed to account for the gamma-ray emission is $W_p\simeq 2\times 10^{50}\,$ergs, which is well consistent with the reach of the usual 10\% of the kinetic energy released in SNRs (Aharonian et al. 2004). Note that if the distance of the SNR and MCs is 3.2\,kpc, the inferred gas density of the MC would be 2.5 times higher (see Table 2), and this change reduces the requirement for the proton energy to $10^{49}\,$ergs.

For the leptonic component, we used an electron/positron broken-power law distribution, i.e., $dN/dE_e\propto E_e^{-s_{e,1}}$ for $E_e<E_b$ and  $dN/dE_e\propto E_e^{-s_{e,2}}$ for $E_e\geq E_b$,  which is usually chosen to describe the SED of PWNe (Tanaka \& Takahara 2010; Bucciantini et al. 2011; Martin et al. 2012; Torres et al. 2013; Torres et al. 2014).
The time-independent spectra of synchrotron radiation and  IC radiation are calculated following Blumenthal (1970).
The IC emissivity is calculated in the optically thin case, which is appropriate for these objects, using the general Klein-Nishina differential cross-section.
We adopt the interstellar radiation field modelled in Popescu et al. (2017) as well as the cosmic microwave background as the target photon field for the IC scattering.
The \fermi\/ data above 10\,GeV together with the HAWC data can be reproduced with $s_{e,1}=1.5$, $s_{e,2}=3.0$ and a break energy at $E_b=8.2\,$TeV, as well as a total energy of $W_e\simeq 4\times 10^{47}\,$erg in the emitting electrons/positrons\footnote{Note that the required total energy is not sensivite to the chosen maximum energy and minimum energy of the spectrum given $s_{e,1}<2$ and $s_{e,2}>2$}.
Assuming the age of the system equal to the characteristic age of \psrj\/, i.e., 20\,kyr, we can roughly evaluate the average magnetic field of a PWN from such a large breaking energy, via equating the age of the system to the cooling timescale of the electrons at the breaking energy $t_{\rm cool}\simeq 40 (E/8.2\rm TeV)^{-1}\left[(U_{\rm ph}+U_B)^{-1}/1\,{\rm eV~cm^{-3}}\right]^{-1}\,$kyr.
Given $U_{\rm ph}=1\,\rm eV~cm^{-3}$ and $U_B=B^2/8\pi$, the inferred magnetic field strength is $B\simeq 6 \mu G$.
Such a magnetic field is consistent with the  X-ray upper limit posed by \xmm.
The low magnetic field strength is similar to some other relic nebulae  in the TeV regime (Aharonian et al. 2006; H.E.S.S. Collaboration 2012; Liu et al. 2019) associated to intermediate-aged pulsars.
Some of those PWNe display energy-dependent morphology in the TeV regime (Aharonian et al. 2006; H.E.S.S. Collaboration 2012; H.E.S.S. Collaboration 2019).
Above 30 GeV, the best fit to the LAT data is the template adopted from VERITAS data, consistent with the morphology obtained in TeV range.
If the emission is indeed related to the pulsar, we expect the nebula to be more compact and closer to the pulsar at TeV energies.
The energy-dependent morphology could be the key to understand the transport mechanism of particles within the PWN and the evolution of the PWN (H.E.S.S. Collaboration 2019; Liu \& Yan 2020).
Deep observations with TeV observatories such H.E.S.S., HAWC or LHAASO will provide crucial input to disentangle the origin of the gamma-ray emission observed.

We also note that there are actually many relevant physical processes which can influence the modelling, such as the particle injection history, particle spectral evolution and particle transport.
We leave such a more realistic modelling to the future study and here we simply test the feasibility of the hybrid interpretation.
In the considered hybrid scenario, the neutrino emission from \mgro\ would not be detectable by current instruments which are operating above 100 GeV (e.g. IceCube). This is because the neutrino spectrum arising from $pp$ collisions generally resemble that of the pionic gamma-ray spectrum, which is important only below $\sim 10\,$GeV and drops quickly with energy. However, we should also note that it is not clear for the time being that whether the gamma-ray spectrum above 100\,TeV would decline as our model expectation. {IceCube Collaboration (Aartsen et al. 2020) found that the best-fit $\nu_\mu+\bar{\nu}_\mu$} number in this region is 4.2 and the best-fit spectral slope is $-$2. This is translated to a 90\% C.L. upper limit for the neutrino flux as $dN/dE_\nu=5.7\times10^{-13}(E_\nu/1\rm TeV)^{-2} TeV^{-1}~cm^{-2}s^{-1}$. In the $pp$ collision, the pionic gamma-ray flux is about twice that of the $\nu_\mu+\bar{\nu}_\mu$ flux. Therefore, an upper limit for the accompanying pionic gamma-ray component can be obtained as shown in the dashed grey line in Figure~\ref{multiSED}. If a hardening of the gamma-ray spectrum beyond 100\,TeV presents in the future observation by HAWC or LHAASO, a { second} hadronic gamma-ray component as well as neutrino detection may then be {confirmed}.

~\\

\acknowledgments

The \textit{Fermi} LAT Collaboration acknowledges generous ongoing support
from a number of agencies and institutes that have supported both the
development and the operation of the LAT as well as scientific data analysis.
These include the National Aeronautics and Space Administration and the
Department of Energy in the United States, the Commissariat \`a l'Energie Atomique
and the Centre National de la Recherche Scientifique / Institut National de Physique
Nucl\'eaire et de Physique des Particules in France, the Agenzia Spaziale Italiana
and the Istituto Nazionale di Fisica Nucleare in Italy, the Ministry of Education,
Culture, Sports, Science and Technology (MEXT), High Energy Accelerator Research
Organization (KEK) and Japan Aerospace Exploration Agency (JAXA) in Japan, and
the K.~A.~Wallenberg Foundation, the Swedish Research Council and the
Swedish National Space Board in Sweden. Additional support for science analysis during the operations phase is gratefully acknowledged from the Istituto Nazionale di Astrofisica in Italy and the Centre National d'\'Etudes Spatiales in France. {This work performed in part under DOE Contract DE-AC02-76SF00515.}

J. L. acknowledges the support from the Alexander von Humboldt Foundation and the National Natural Science Foundation of China via NSFC-11733009.
R.-Y. L. acknowledges the support from the National Natural Science Foundation of China via NSFC-U2031105.
E. O. W. acknowledges the support from the Alexander von Humboldt Foundation.
The work of D. F. T. has been supported by the grants PGC2018-095512-B-I00, SGR2017-1383, and AYA2017- 92402-EXP.
Q. C. L. acknowledges support from the program A for Outstanding PhD candidate of Nanjing University.
Work at NRL is supported by NASA.

\textbf{Appendix A: CO data analysis }\\

\label{MC}

The CO data used in this work are part of the Milky Way Imaging Scroll Painting (MWISP) project (Su et al. 2019), including \twCO\ (\otz), \thCO\ (\otz), and \CeiO\ (\otz) which were observed simultaneously using the 13.7\,m millimeter-wavelength telescope of the Purple Mountain Observatory at Delingha.
A new $3\times3$ pixel Superconducting Spectro-scopic Array Receiver was used as the front end (Shan et al. 2012).
The bandwidth of the spectrometer is 1\,GHz and the half-power beamwidth of the telescope is about $50''$ at 115\,GHz.
The spectral resolution is 61\,KHz, corresponding to velocity resolutions of $0.16\km\ps$ for \twCO\ and $0.17\km\ps$ for \thCO\ and \CeiO.
The typical rms noise level is about 0.5\,K for \twCO\ (\otz) and 0.3\,K for \thCO\ (\otz) and \CeiO\ (\otz).

\begin{figure}
\centering

\includegraphics[scale=0.4]{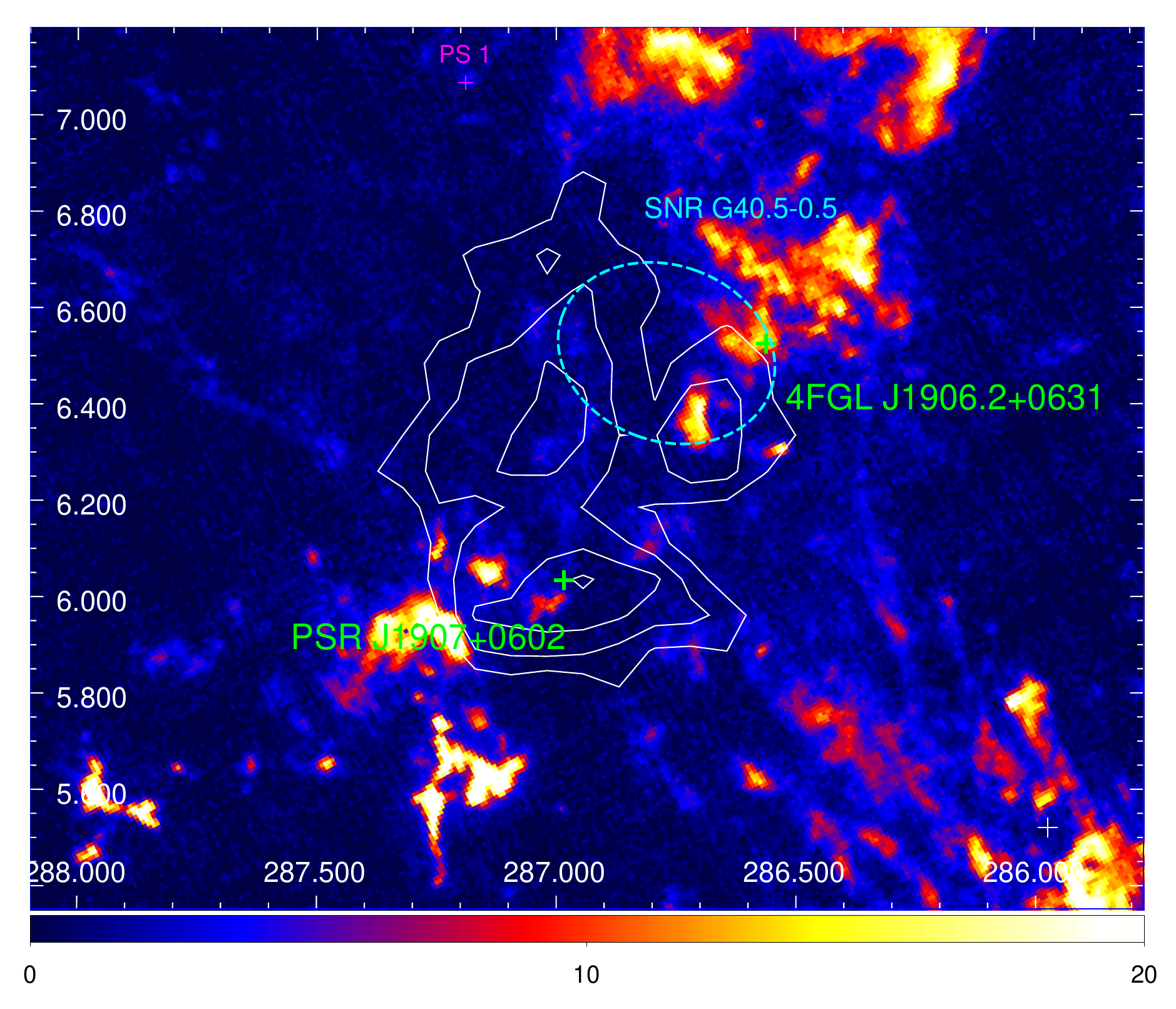}
\includegraphics[scale=0.4]{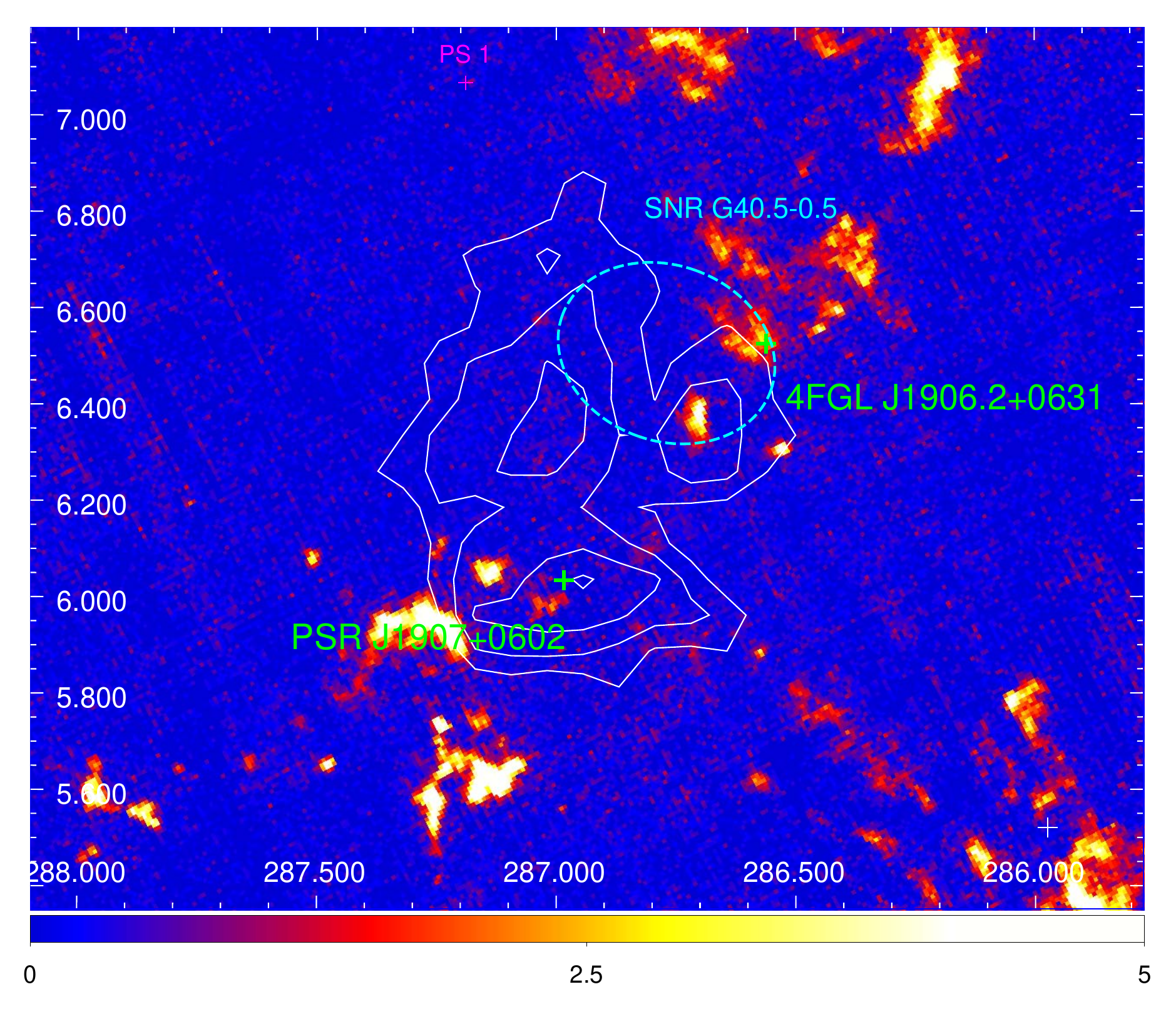}\\
\includegraphics[scale=0.4]{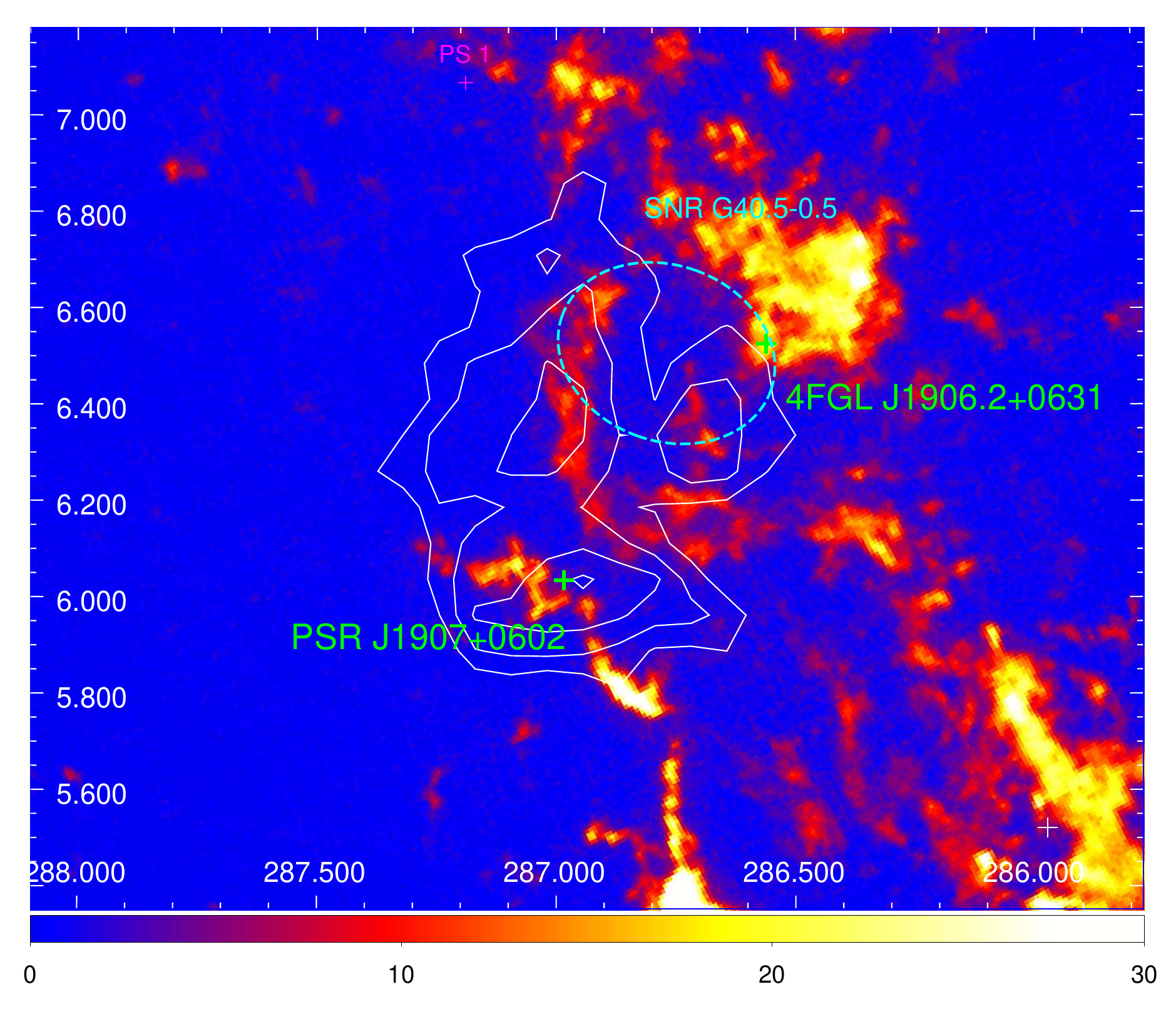}
\includegraphics[scale=0.4]{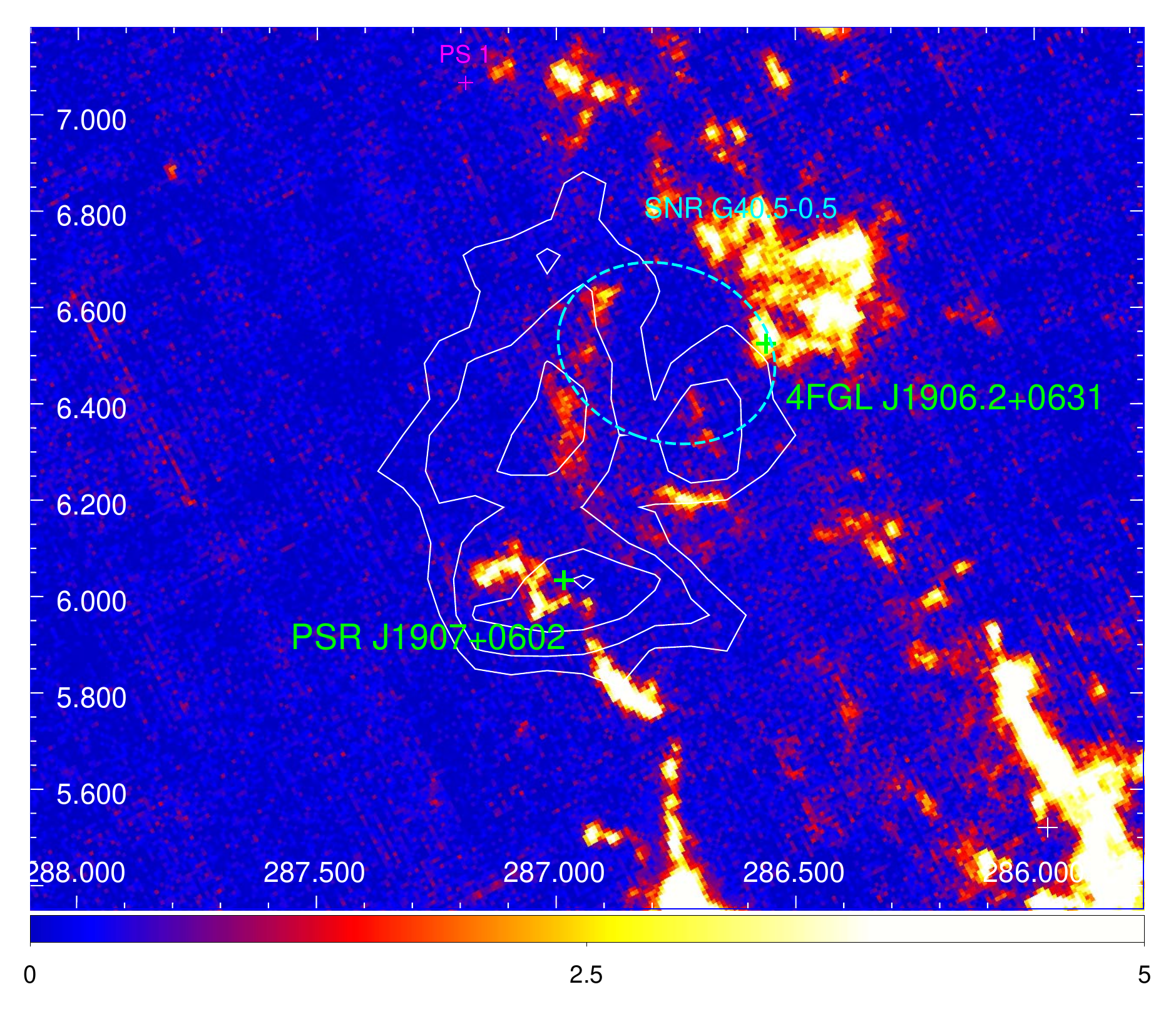}\\
\caption{From top to bottom: \twCO\ (\otz) (left) and \thCO\ (\otz) (right) intensity map (in the unit of K $\km\ps$) integrated in velocity range 46 -- $50\km\ps$, 50 -- $54\km\ps$, 54 -- $58\km\ps$,
58 -- $62\km\ps$, 62 -- $66\km\ps$.
The color denotes the intensity.
White contours correspond to the VERITAS significance map starting from 3$\sigma$ with a step of 1$\sigma$.
The labels are as in Figure \ref{tsmap}.
The x and y axes are R.A. and decl. (J2000) in degrees.
\label{COvelosities}}
\end{figure}

MCs are observed to be spatially associated with the TeV emission (Figure \ref{co}).
Figure \ref{COvelosities} displays the \twCO\ (\otz) and \thCO\ (\otz) maps for five consecutive velocity ranges from 46 $\km\ps$ to 66 $\km\ps$, with a coverage of 4$\km\ps$.

\begin{figure}
\centering
\includegraphics[scale=0.805]{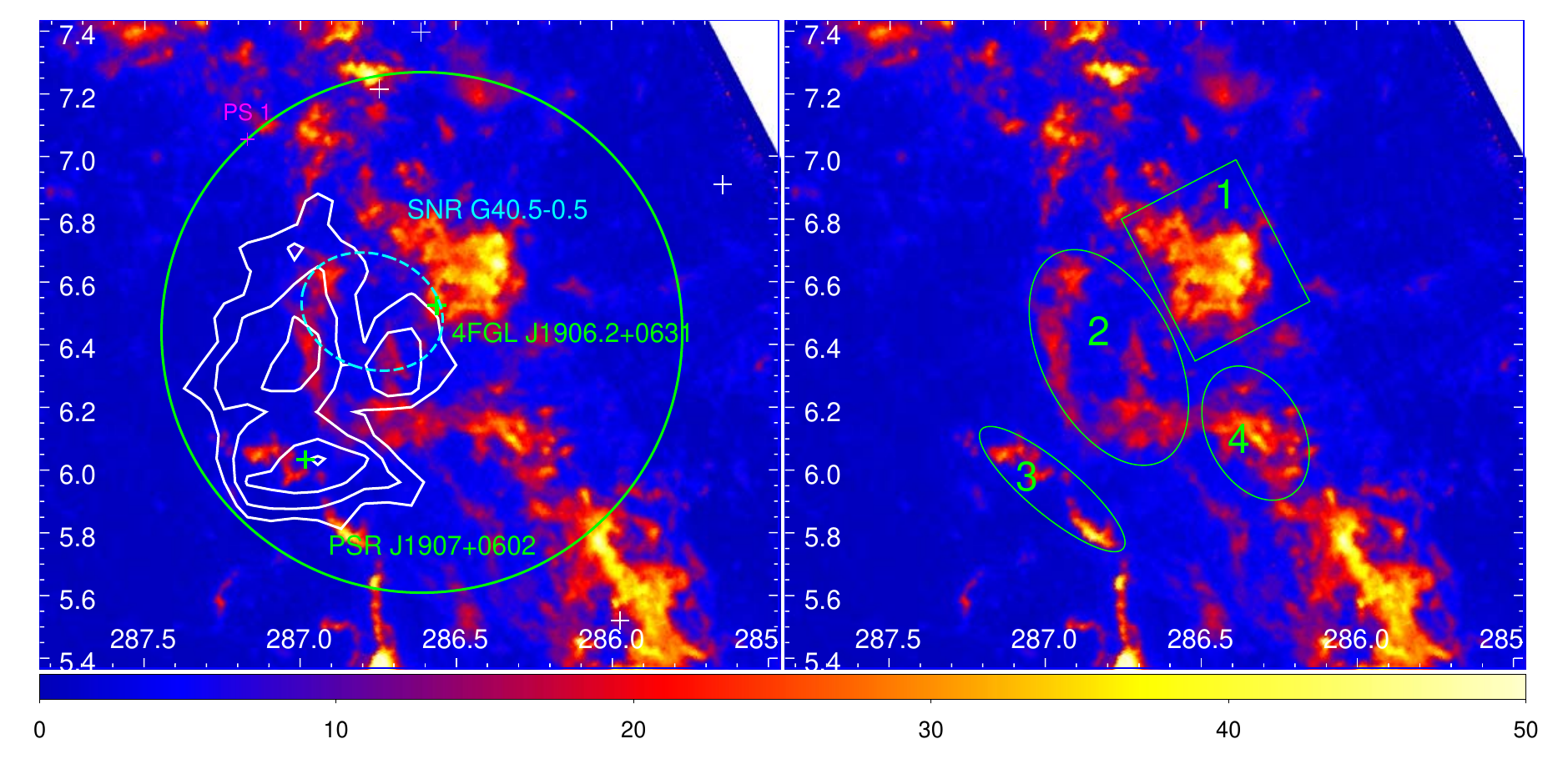}
\includegraphics[scale=0.4]{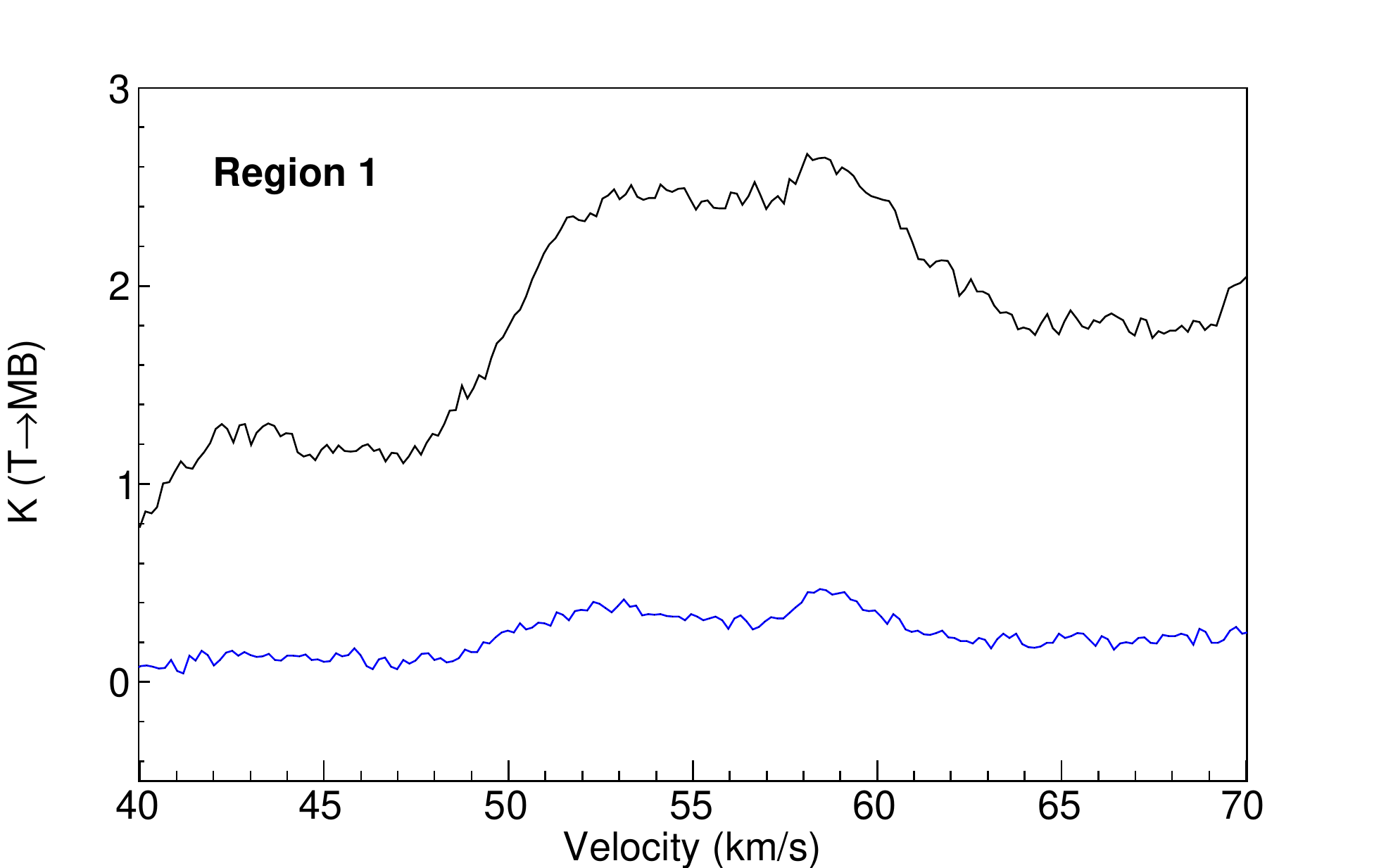}
\includegraphics[scale=0.4]{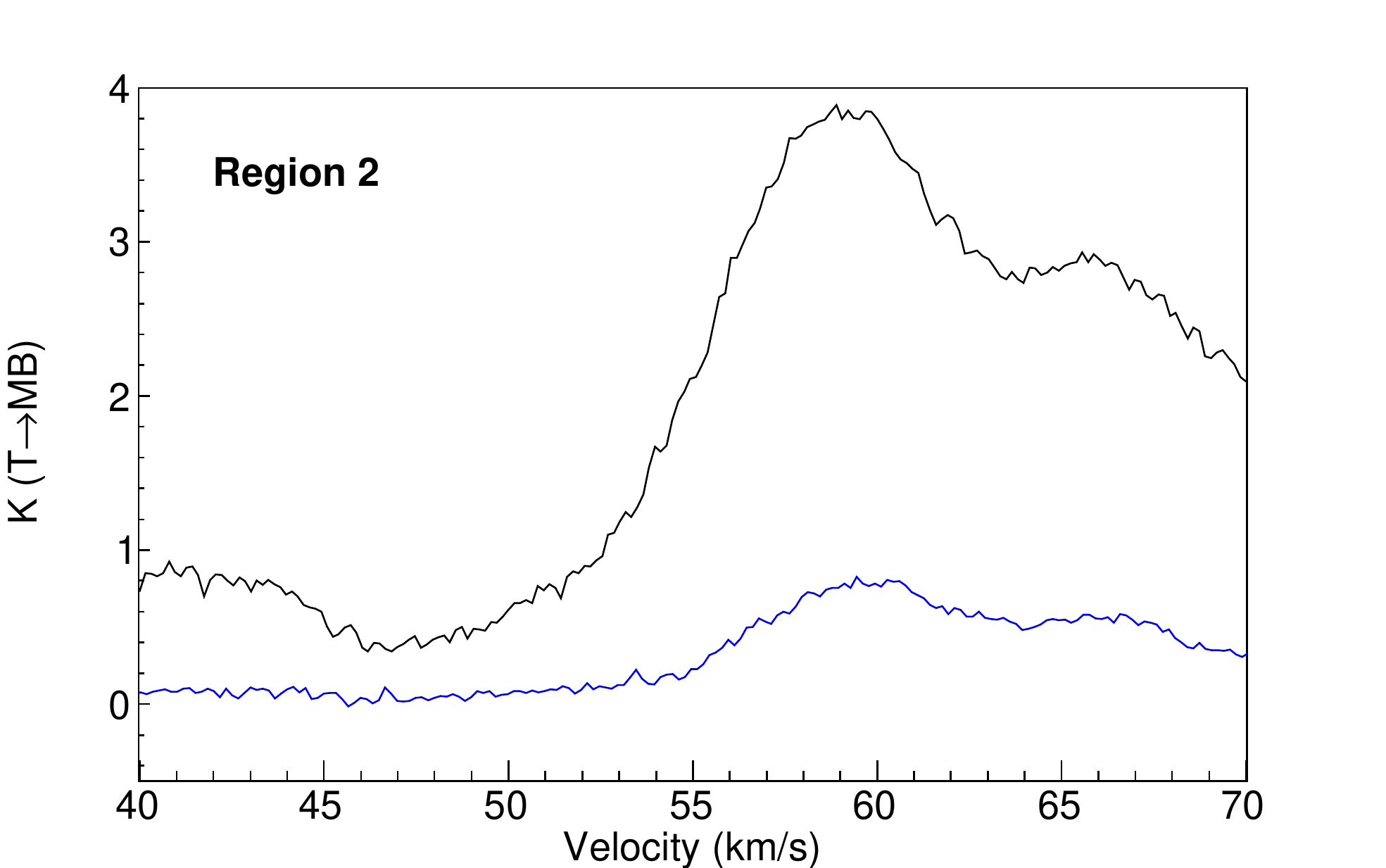}\\
\includegraphics[scale=0.4]{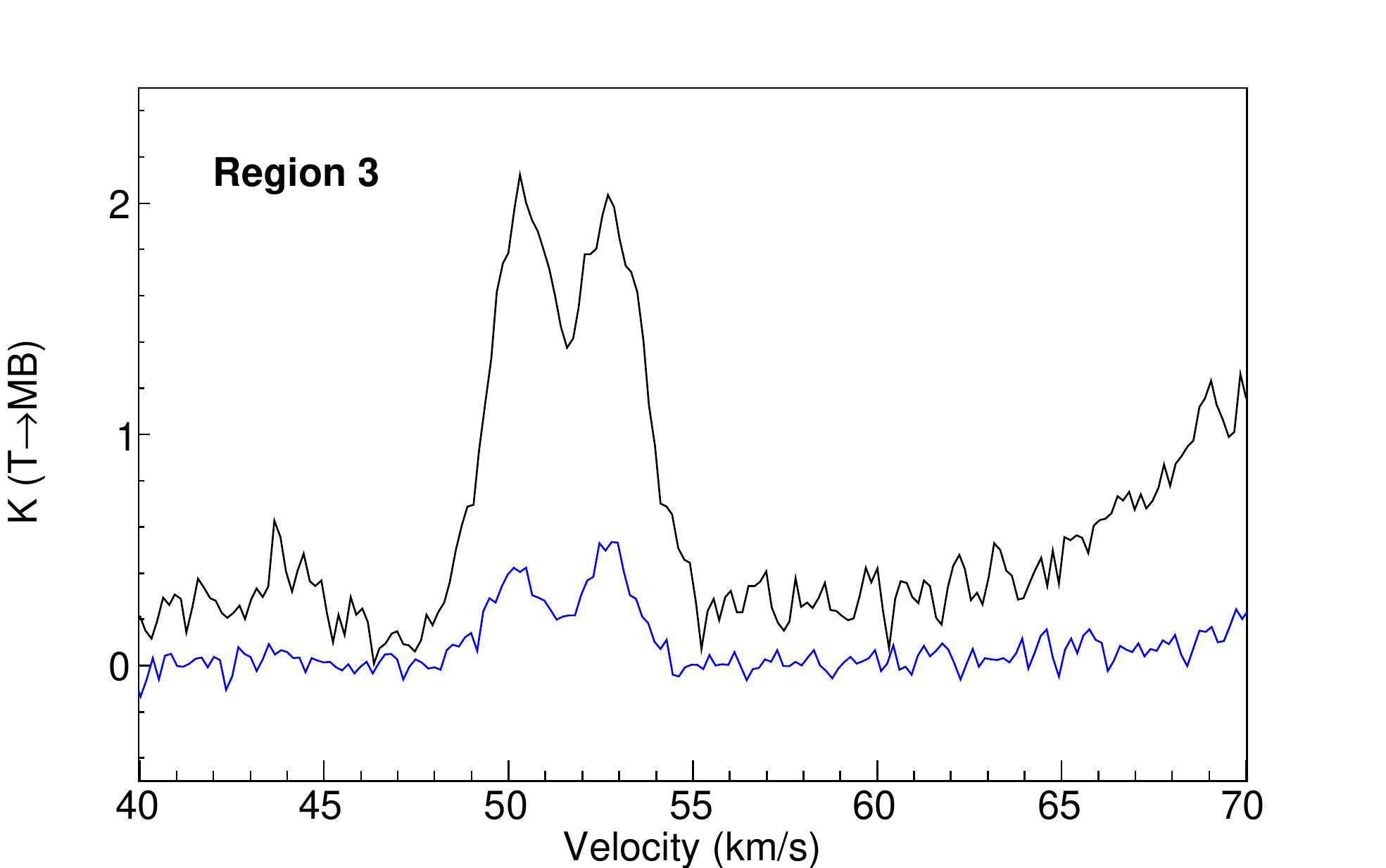}
\includegraphics[scale=0.4]{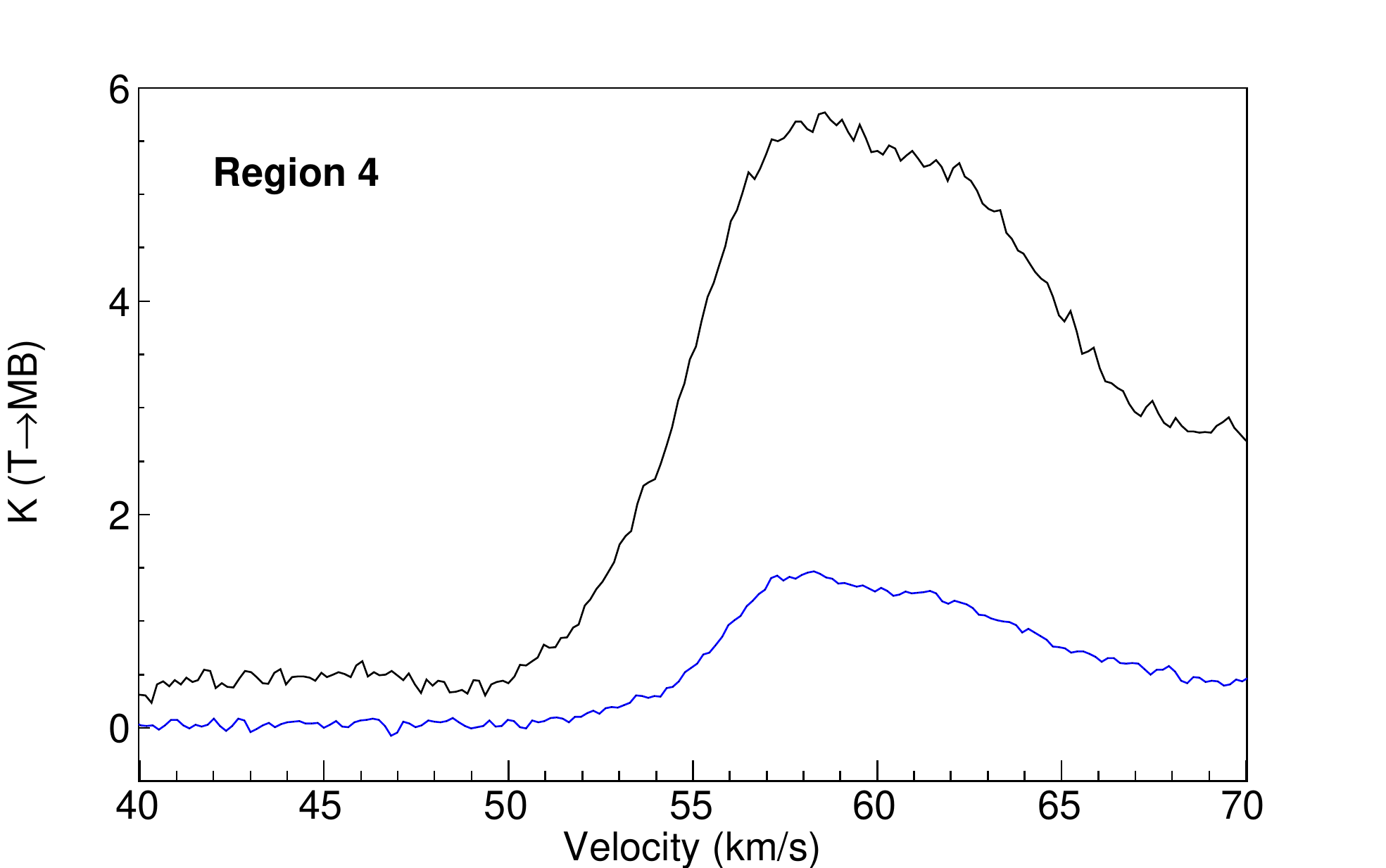}

\caption{\textbf{Top panel}: \twCO\ (\otz) intensity map (in the unit of K $\km\ps$) integrated in velocity range 49 -- $56\km\ps$, shown twice.
White contours correspond to VERITAS significance map (Aliu et al. 2014) starting from 3$\sigma$ to 6$\sigma$ by 1$\sigma$ steps.
The green circle shows the disk morphology of Fermi J1908+06 (see Section 3.2).
The x and y axes are R.A. and decl. (J2000) in degrees.
Four regions delineated in green and labelled with roman numerals ``1'' to ``4'' are used to estimate the astrophysical parameters for
the molecular gases (see Table~\ref{COparameter}).
\textbf{Lower panels}: \twCO\ (\otz\, black) and \thCO\ (\otz\/, blue) spectra of region ``1'' to ``4''.}

\label{COregion}
\end{figure}

 We made an estimation of the astrophysical properties of the MCs in 4 regions and have parameterized the distance as $d=7.9d_{7.9}\kpc$.
We estimated the kinematic distance to the MCs using the Milky Way's rotation curve suggested by Brand \& Blitz (1993), assuming
the Sun's Galactocentric distance to be 8.5\,kpc and orbital speed to be $220\,{\rm km}\,{\rm s}^{-1}$.
Therefore, the velocity of each MC could indicate two candidate kinematic distances, the near side one and the far side one.
\twCO (\otz) and \thCO (\otz) spectra of the 4 regions are extracted and shown in Figure \ref{COregion}.
The estimated results are summarized in Table~\ref{COparameter}.
Among the parameters, the column densities are estimated by using \thCO\ lines, under the assumption of local-thermal-equilibrium (LTE) condition, optically thin conditions
for \thCO\ (\otz) line, and optically thick conditions for \twCO\ (\otz).
The excitation temperature is assumed to be $T_{\rm ex} = 12$\,K, 12\,K, 21\,K, and 15\,K, respectively, the value estimated from the maximum \twCO\ (\otz) line emission.
Also, the conversion relation for the molecular column density of $\NHH \approx 7\times 10^5 N$(\thCO) (Frerking et al. 1982) has been used.
The mean density of each region is calculated by dividing the column density toward the \thCO\ emission peaks by the cloud size along the line-of-sight,
which is estimated from the FWHM of the \thCO\ line with Larson's law (Larson 1981).
Furthermore, a mean density of the four regions, when taking the weight of the volume into consideration, is estimated to be $\sim$ 45 ${\rm cm}^{-3}$.

\begin{figure}
\centering

\includegraphics[scale=0.4]{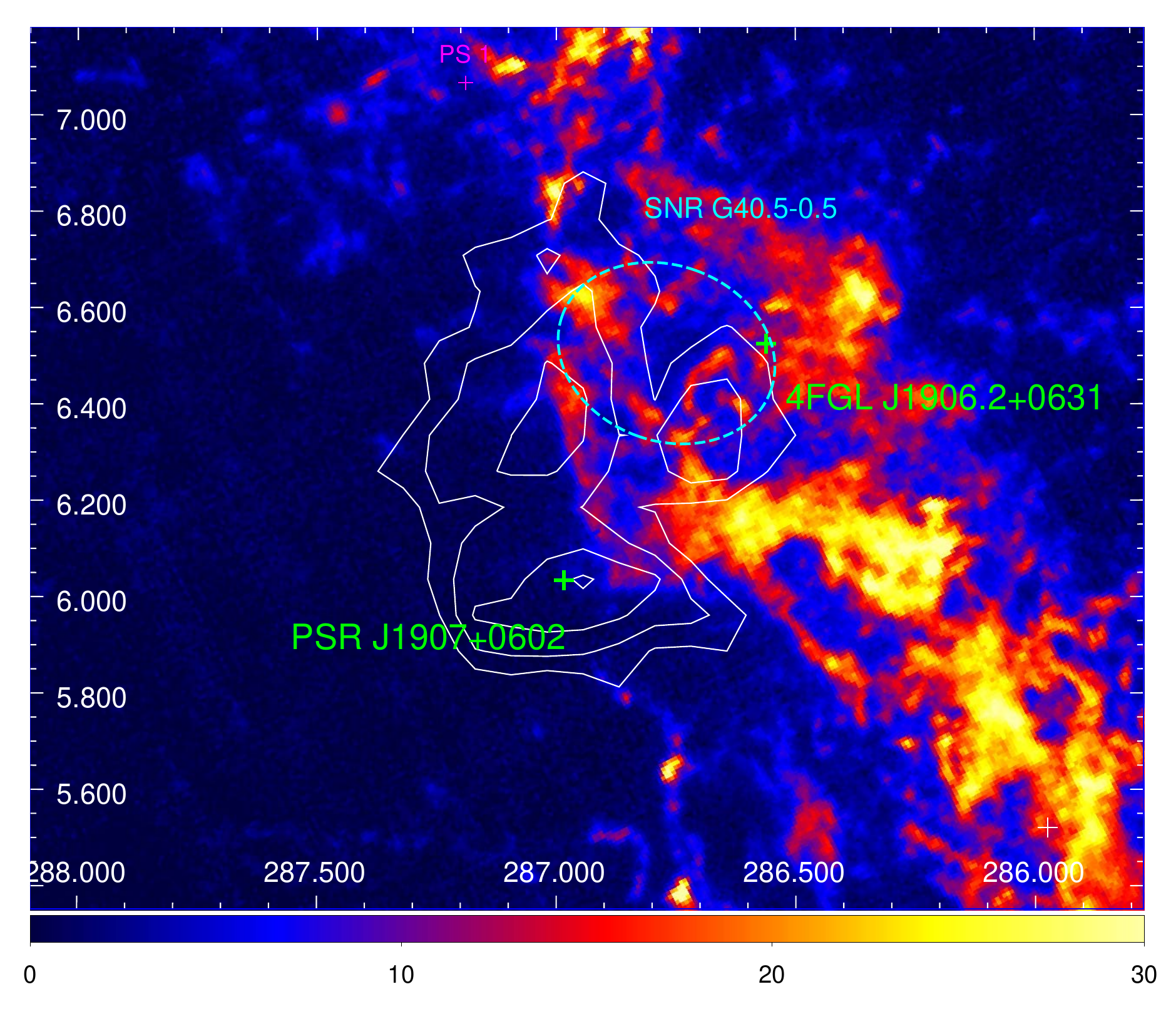}
\includegraphics[scale=0.4]{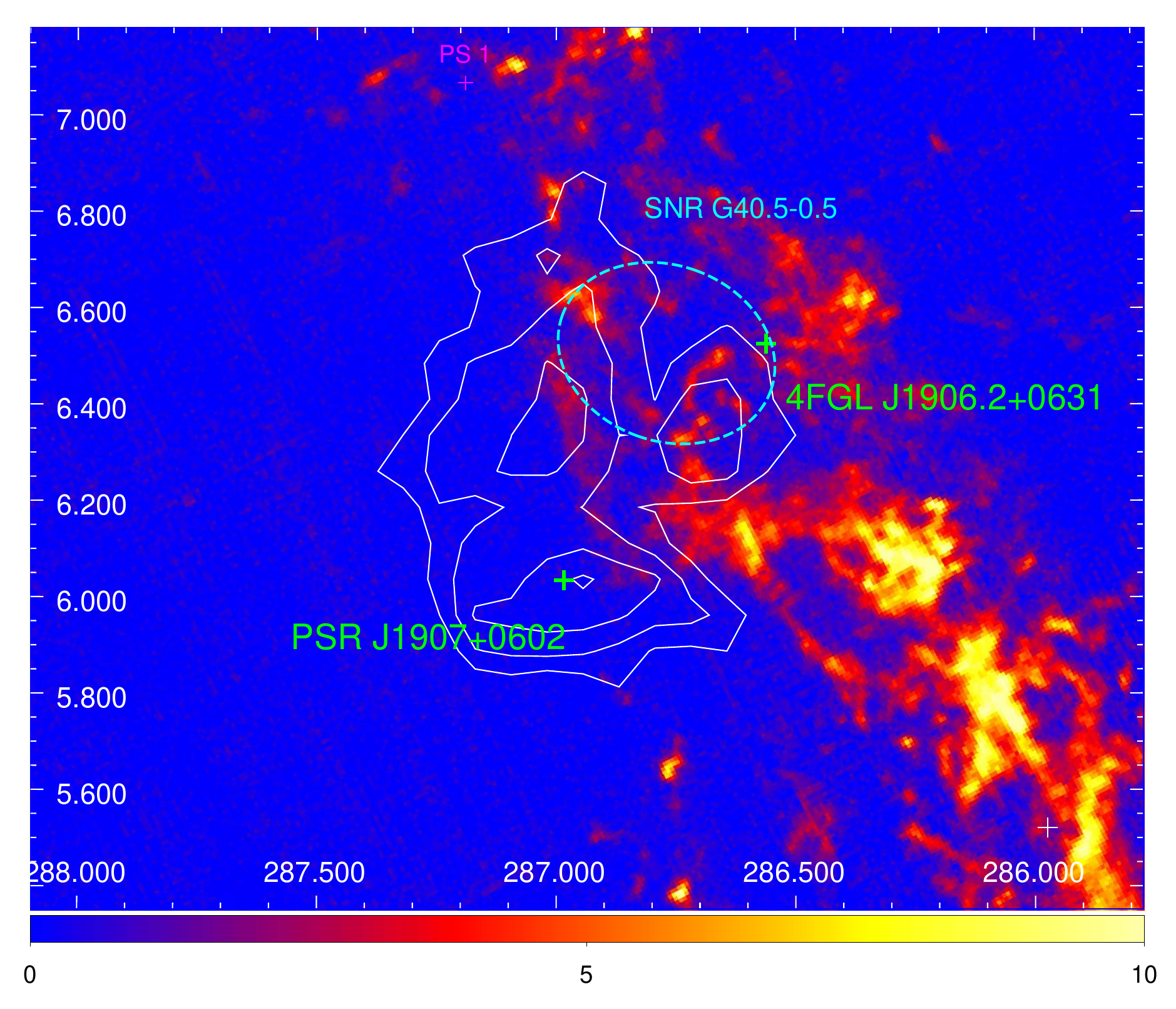}\\
\includegraphics[scale=0.4]{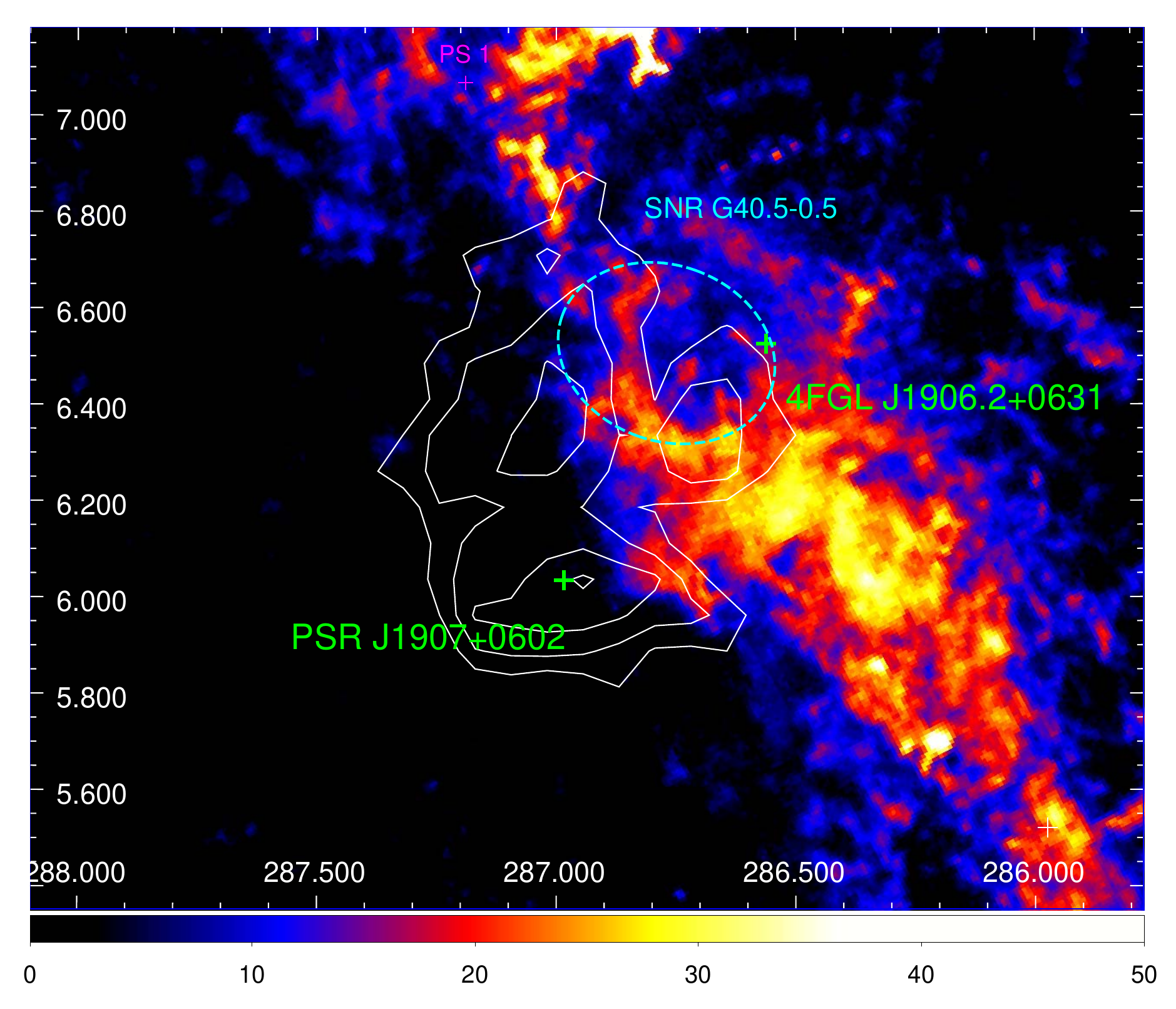}
\includegraphics[scale=0.4]{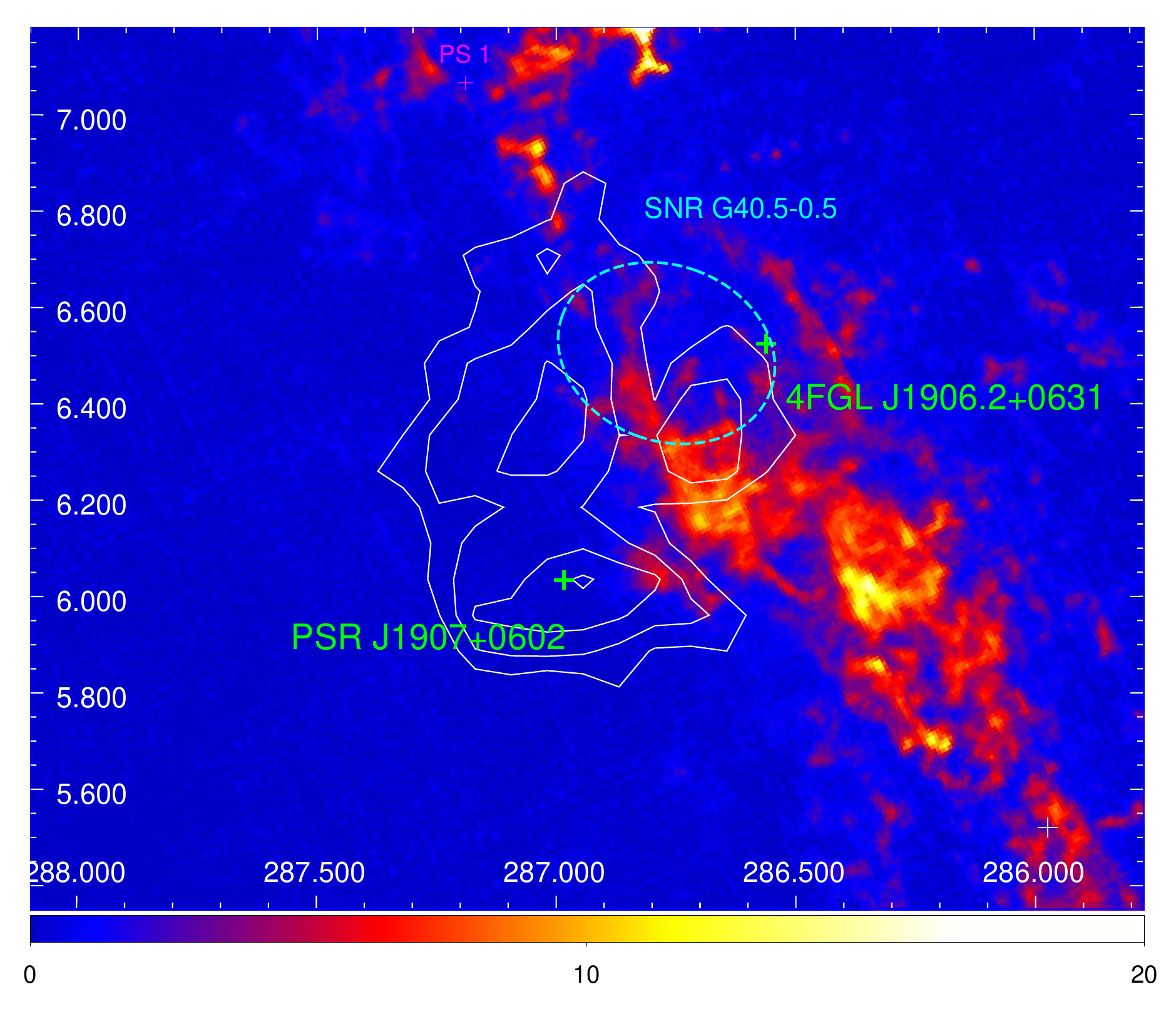}\\
\includegraphics[scale=0.4]{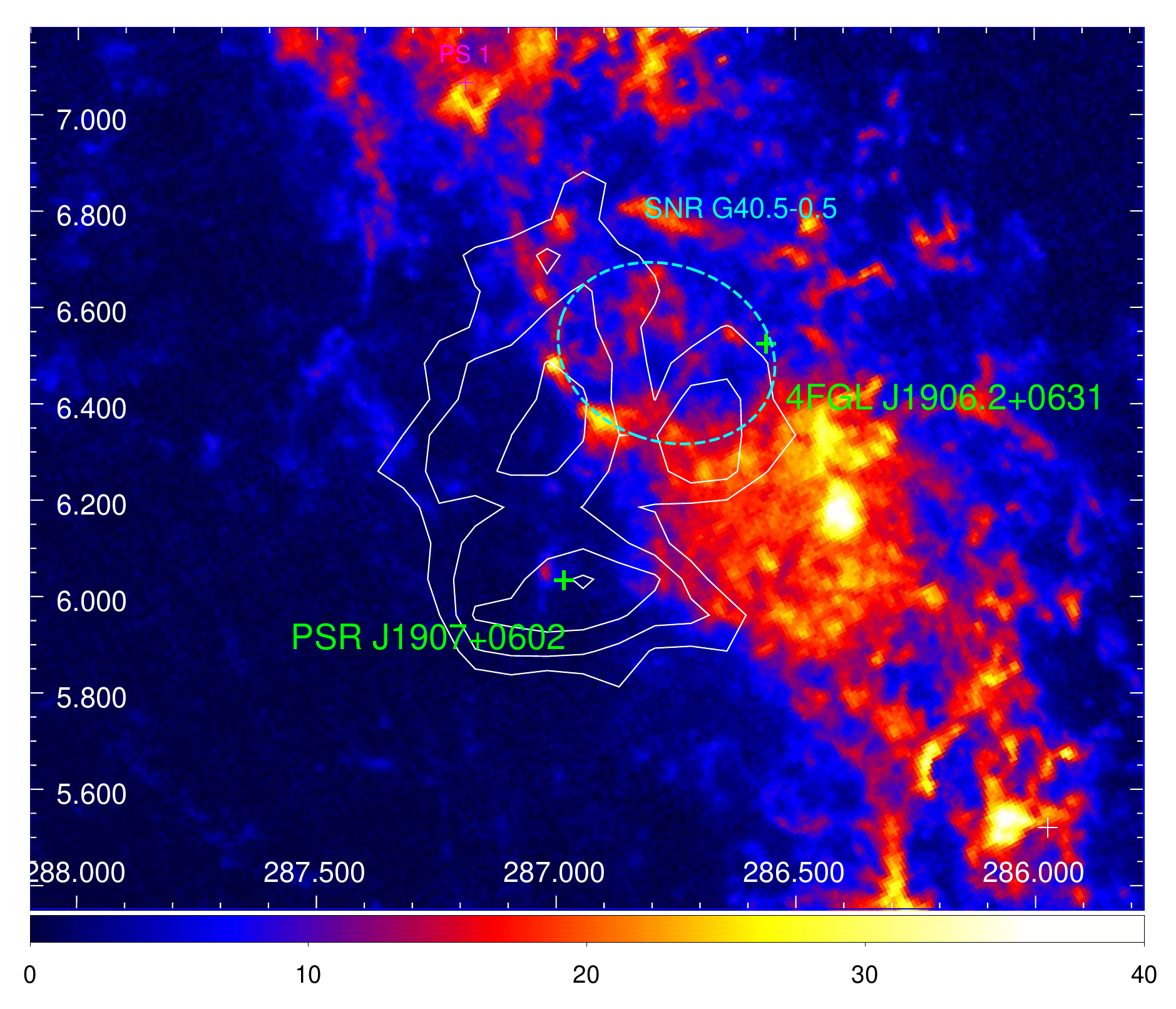}
\includegraphics[scale=0.4]{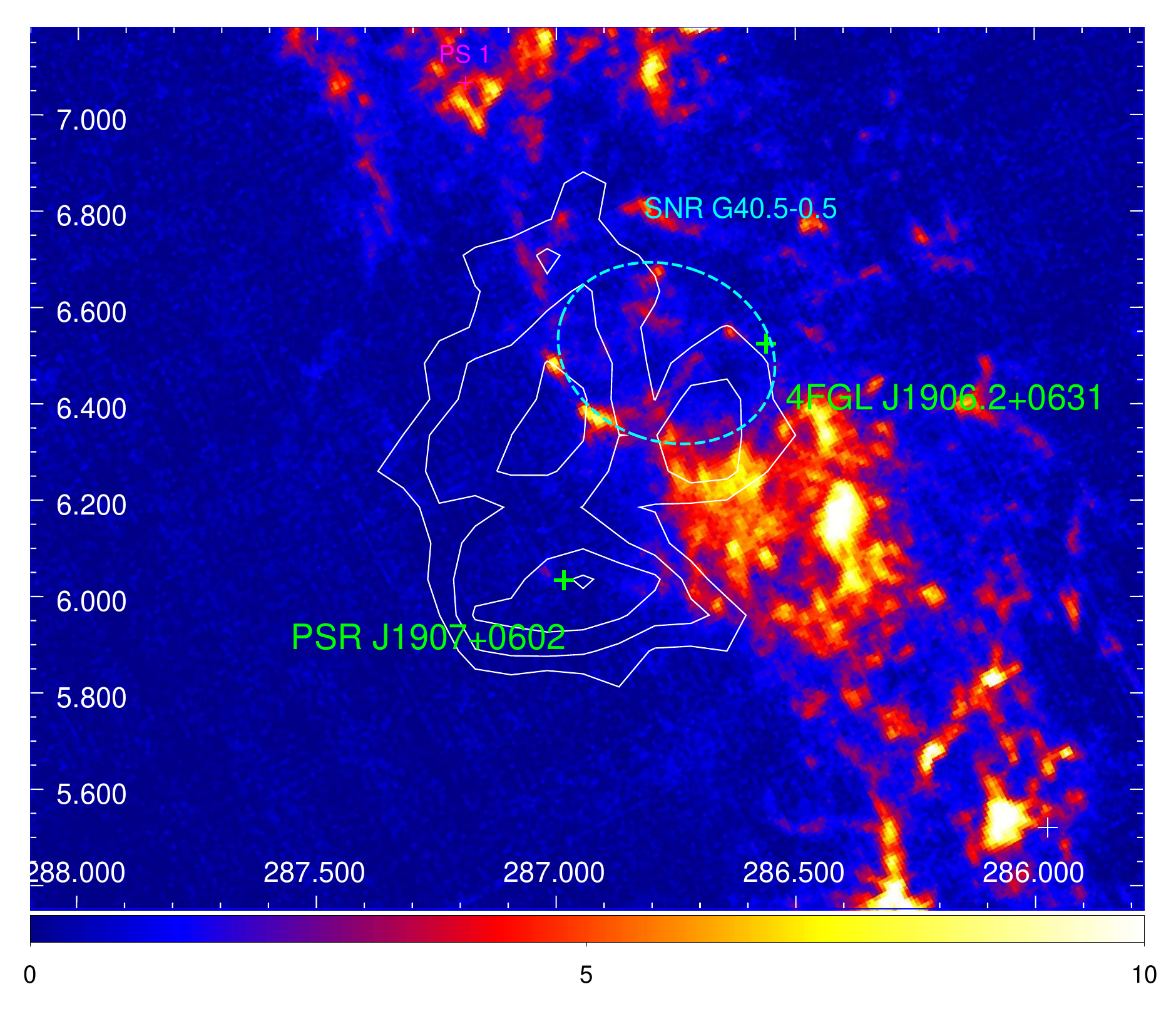}\\

-Figure \ref{COvelosities} continues

\end{figure}

\begin{table*}{}
\centering
\caption{Fitted and Derived Parameters for the MCs around $50\km\ps$ in 4 regions as indicated in Figure \ref{COregion}}

\begin{tabular}{ccccccc}
\\
\\
\hline\hline

Region & $N$(H$_2$) & $n(\mbox{H}_2)$  & $M(\mbox{H}_2)$    & FWHM & Line Center & Near/Far Distance\\
            & ($10^{21}$cm$^{-2}$)             & (cm$^{-3})$                      & $(10^3 M_\odot)$ &  ($\km\ps$)   & ($\km\ps$) & (kpc) \\

\hline\hline                 

1 & 1.2 & 16$d_{7.9}^{-1}$ & 55.6$d_{7.9}^2$ & 2.7 & 52.9 & 3.6/9.3 \\
2 & 3.2 & 16$d_{7.9}^{-1}$ & 166.9$d_{7.9}^2$ & 3.8 & 57.9& 4.0/8.9\\
3 & 1.0 & 180$d_{7.9}^{-1}$ & 23.1$d_{7.9}^2$ & 1.7 & 50.1 &3.4/9.5\\
4 & 3.4 & 24$d_{7.9}^{-1}$ & 125.0$d_{7.9}^2$ & 3.5 & 56.9 &3.9/9.0\\

\hline\hline                 
\\
\label{COparameter}
\end{tabular}
\end{table*}

\textbf{Appendix B: {\it Fermi}-LAT data analysis}\\


The analysis shown in this paper uses more than 11 years of \emph{Fermi}-LAT P8R3 data, from 2008 August 4 (MJD 54682) to 2019 November 09 (MJD 58796).
All gamma-ray photons within a circular region of interest (ROI) of 15$\degree$ radius centered on \psrj\/ are considered in our analysis.
The gamma-ray events from ``Pass 8'' event reconstruction are analyzed using the \emph{Fermi} Science Tools 11-07-00 release.
In the data reduction, a zenith angle threshold of 90$\degree$ is adopted to reject contamination from gamma rays from the Earth's limb.
The selected \emph{Fermi}-LAT instrument response functions (IRFs) is ``P8R3 V2 Source''.
Known gamma-ray sources from the \emph{Fermi} Large Area Telescope Fourth Source Catalog (4FGL, Abdollahi et al. 2020) within $20\degree$ of \psrj\/ were included in the spectral-spatial model, as well as Galactic (``gll\_iem\_v07.fits") and isotropic diffuse emission components  (``iso\_P8R3\_SOURCE\_V2\_v1.txt").
The spectral parameters of the sources within 4$\degree $ of \psrj\/, Galactic and isotropic diffuse emission components were all left free.
\psrj\/ and \fgl\/ are spatially associated with \mgro\/, thus not included in the spectral-spatial model.
The spectral parameters of sources with larger angular separations were fixed at the 4FGL values.
The spectral analysis was performed using a binned maximum likelihood fit (spatial bin size 0.1$\degree$ and 30 logarithmically spaced bins in the 0.1--300 GeV range)
For the analysis in 30 GeV--1 TeV,  the spectral parameters of the sources within 4$\degree $ of \psrj\/ and the Galactic diffuse emission component are fixed to 4FGL values except for the prefactor (spectral normalization) because of low statistics.

The significance of the sources was evaluated by the Test Statistic (TS).
This statistic is defined as TS=$-2 \ln (L_{max, 0}/L_{max, 1})$, where $L_{max, 0}$ is the maximum likelihood value for a model in which the source studied is removed (the ``null hypothesis"), and $L_{max, 1}$ is the corresponding maximum likelihood value with this source being incorporated.
The square root of the TS is approximately equal to the detection significance of a given source.
The significance of source extension was defined as {TS$_{ext}$=$-2\ln (L_{point}/L_{ext}$)}, where $L_{ext}$ and $L_{point}$ are
the \textit{gtlike} global likelihood of the extended source hypotheses and the point source, respectively.
The threshold for claiming the source to be spatially extended is set as TS$_{ext}>$16, {which corresponds} to a significance of $\sim$ 4$\sigma$.
The source localization, extension fitting and TS maps production were carried out using the \textit{Fermipy} analysis package (version 0.17.4; Wood et al. 2017).
Energy dispersion correction has been applied in the analysis.
The SEDs are computed assuming a power-law shape with spectral index fixed at 2.

\begin{figure*}
\centering
\includegraphics[scale=0.7]{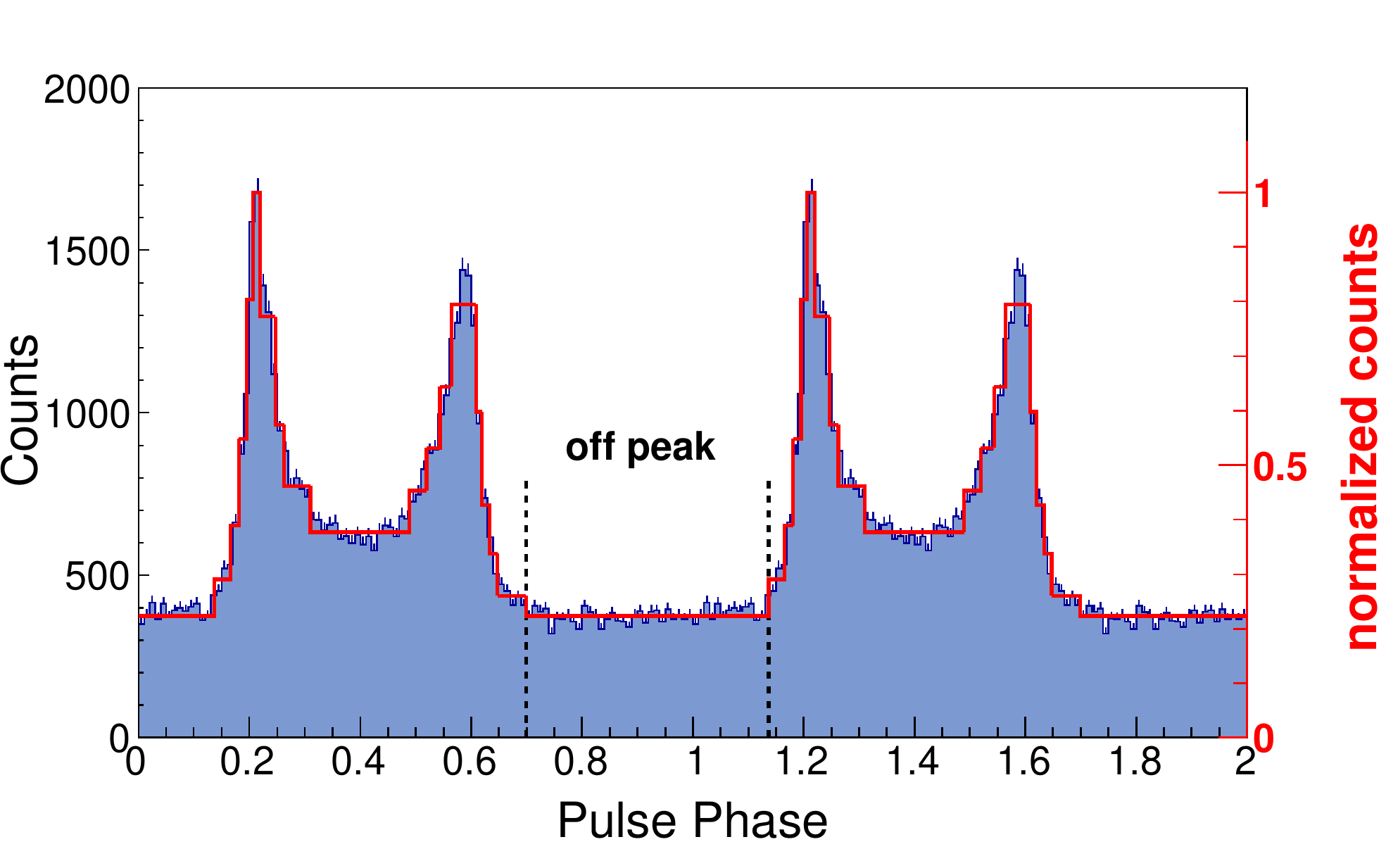}

\caption{Pulse profile of \psrj\/ above 300 MeV with an ROI of 0.6$\degree$.
Two rotational pulse periods are shown, with a resolution of 100 phase bins per period.
The Bayesian block decomposition from Li et al. (2020) is shown by red lines.
The off-peak intervals ($\phi$=0.0$-$0.136 and 0.697$-$1.0) are defined by black dotted lines.}
\label{profile}
\end{figure*}

\psrj\/ is a bright gamma-ray pulsar spatially associated with \mgro\/.
To minimize contamination from this pulsar, we carried out data analysis during the off-peak phases of \psrj\/.
Using \emph{Tempo2} (Hobbs et al. 2006) with the {\it Fermi} plug-in (Ray et al. 2011), we have assigned pulsar rotational phases for each gamma-ray photon that passed the selection criteria, adopting the most updated ephemeris for \psrj\/.
The pulse profile of \psrj\/ is shown in Figure \ref{profile}.
We followed the off-peak definition in Li et al. (2020), which is $\phi$=0.0$-$0.136 and 0.697$-$1.0.
To account for the off-peak phase selection, the prefactor parameter of all sources were scaled by 0.439, the width of the off-peak interval.

In the off-peak analysis of the 0.1--300 GeV band, we searched for significant TS excess beyond \nsource\ in the TS map within 4 degrees of \psrj\/.
A new point source (PS 1 hereafter) is located at  R.A.=287.19\degree $\pm$0.06\degree, decl.=7.07\degree $\pm$0.03\degree (Figure \ref{tsmap}).
Assuming a power-law spectral shape, the likelihood analysis of PS 1 resulted in a TS value of 31, spectral index of {2.27}$\pm$0.10, and an energy flux of (1.77$\pm$0.40) $\times$ 10$^{-11}$ erg~cm$^{-2}$s$^{-1}$.

Since VERITAS observations have the deepest exposure on \mgro\ in TeV range and provided most detailed TeV morphology, we included the VERITAS counts map as a template in our GeV morphology analysis.
HAWC observations provided the only SED data points on \mgro\ above 50 TeV.
Thus HAWC data are adopted in the multi-wavelength SED modelling.
We noticed that in $\sim$1-10 TeV HAWC data points have higher flux than H.E.S.S. (Figure \ref{multiSED}).
This may due to that, Imaging Atmospheric Cherenkov Telescopes (e.g. VERITAS, H.E.S.S.) use blank sky region near gamma-ray sources as background.
In case of a large source extension (e.g. \mgro), there might be dim emissions in the rim taken as background, leading to a lower source flux level.


\begin{figure*}
\centering
\includegraphics[angle=90,scale=0.335]{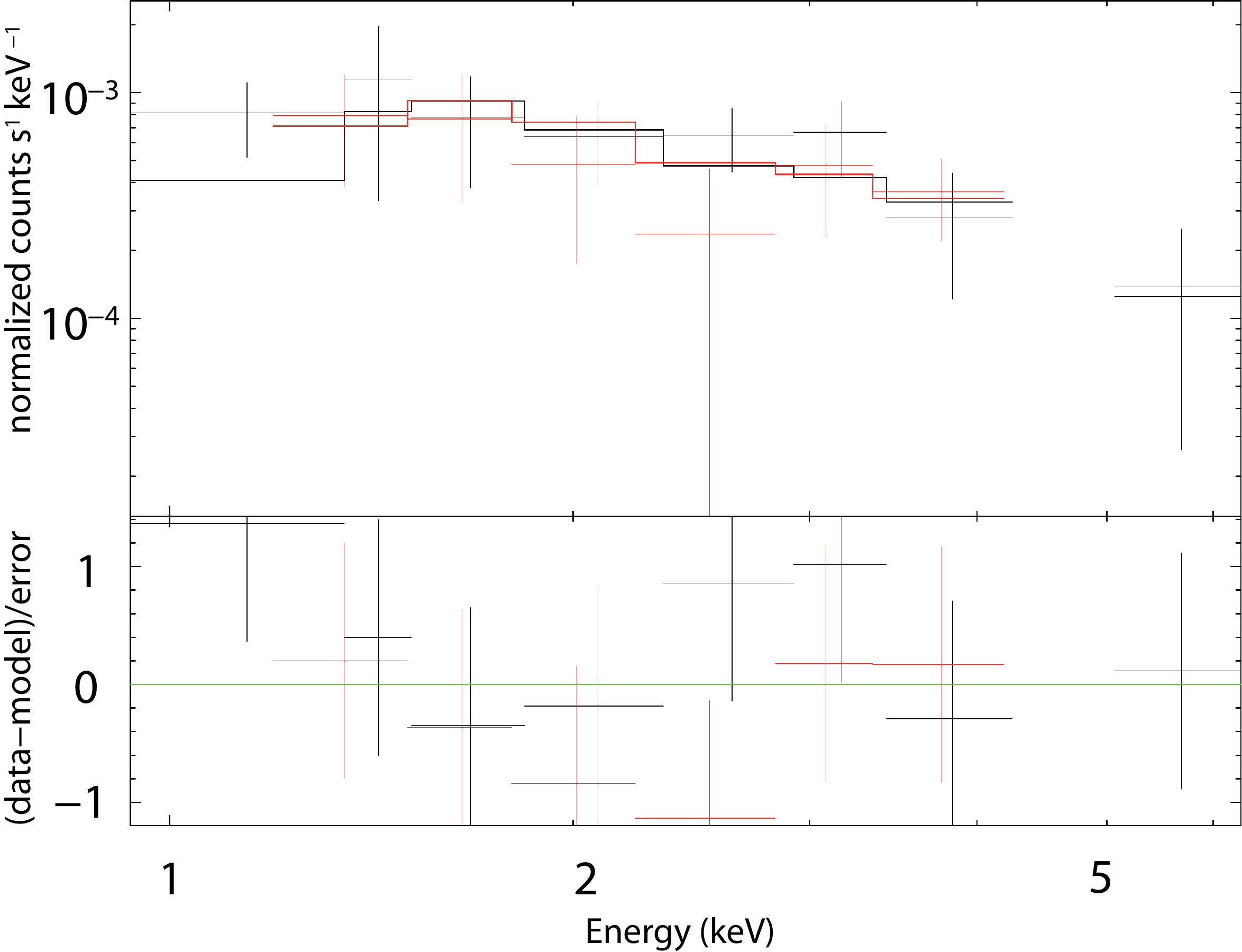}
\includegraphics[angle=90,scale=0.34]{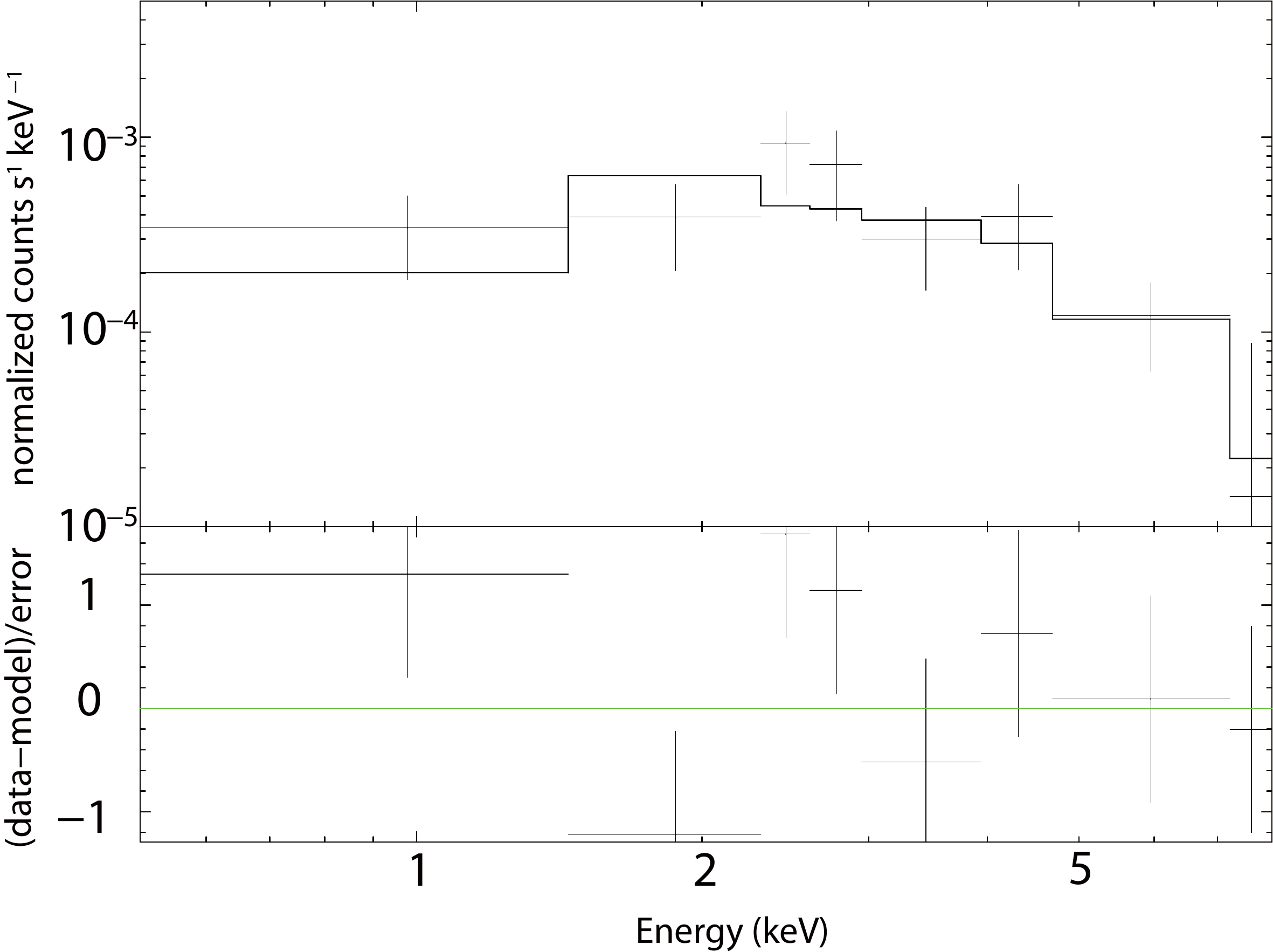}

\caption{\psrj\ \xmm\ MOS 1 (black) and MOS 2 (red) spectra (left) and \emph{Chandra} ACIS spectrum.
The best-fitted models and post-fit residuals are also shown in each case.}
\label{Xray}
\end{figure*}


\textbf{Appendix C: X-ray data analysis}
\label{X-ray}

{\it XMM-Newton} data sets were reduced with the Science Analysis System (SAS, version 16.1.0).
Standard pipeline tasks \emph{emproc} for MOS data were used to process the raw observation data files (ODFs).
{\it XMM-Newton} data were also filtered to avoid the periods of hard X-ray background flares.

The \emph{Chandra} data were reduced using CIAO version 4.7 and CALDB version 4.7.7. We reprocessed the \emph{Chandra} data to level=2 and removed periods of high background or flaring appearing in the observations.

\end{document}